\documentclass[aps,prd,twocolumn,superscriptaddress]{revtex4}
\usepackage{epsfig,epsf}
\usepackage{amsmath}
\usepackage{amsthm}
\usepackage{amsfonts}
\usepackage{amssymb}
\usepackage{dsfont}
\usepackage{multirow}
\usepackage{appendix}
\usepackage{slashed}
\usepackage[active]{srcltx}
\usepackage{psfrag}

\begin{document}

\title{Hadronic tensor molecule $B_c^{\ast +}B_c^{\ast -}$}
\date{\today}
\author{S.~S.~Agaev}
\affiliation{Institute for Physical Problems, Baku State University, Az--1148 Baku,
Azerbaijan}
\author{K.~Azizi}
\affiliation{Department of Physics, University of Tehran, North Karegar Avenue, Tehran
14395-547, Iran}
\affiliation{Department of Physics, Dogus University, Dudullu-\"{U}mraniye, 34775
Istanbul, T\"{u}rkiye}
\author{H.~Sundu}
\affiliation{Department of Physics Engineering, Istanbul Medeniyet University, 34700
Istanbul, T\"{u}rkiye}

\begin{abstract}
Mass and full decay width of the hadronic tensor molecule $\mathcal{M}_{
\mathrm{T}}=B_c^{\ast +}B_c^{\ast -}$ with spin-parities $J^{\mathrm{PC}%
}=2^{++}$ are examined in the framework of QCD sum rule method. To find the
mass $m$ and current coupling $\Lambda$ of this state, we use the two-point
sum rules. The result $(12.87 \pm 0.08)~\mathrm{\ GeV}$ for $m$ indicates
that $\mathcal{M}_{\mathrm{T}}$ can easily decay to pairs of $J/\psi
\Upsilon $, $\eta_{b}\eta _{c}$, $B_{c}^{+}B_{c}^{-}$, and $B_{c}^{\ast
+}B_{c}^{\ast -}$ mesons. Kinematically permitted ways for $\mathcal{M}_{%
\mathrm{T}}$ to transform to conventional mesons include also processes $%
\mathcal{M}_{\mathrm{T}} \to D^{(\ast)+}D^{(\ast )-}$, $D^{(\ast )0}%
\overline{D}^{(\ast )0}$, and $D_{s}^{(\ast)+}D_{s}^{(\ast)-}$ triggered by
annihilation of $b\overline{b}$ quarks in the molecule. The decays to meson
pairs $B^{(\ast)+}B^{(\ast )-}$, $B^{(\ast )0}\overline{B}^{(\ast )0}$ , and
$B_{s}^{(\ast)0}\overline{B}_{s}^{(\ast )0}$ generated by $c\overline{c }
\to $ light quarks are also among possible channels of the hadronic molecule
$\mathcal{M}_{\mathrm{T}}$. Technical tools of the three-point sum rule
approach are applied to compute partial widths of these decays. Our results
for the mass and width $\Gamma_{\mathcal{M}_{\mathrm{T}}}=(154 \pm 19)~
\mathrm{MeV} $ of the tensor molecule $\mathcal{M}_{\mathrm{T}}$ are
important for experimental studies of fully heavy exotic structures.
\end{abstract}

\maketitle

%%%%%%%%%%%%%%%%%%%%%%%%%%%%%%%%%%%%%%%%%%%%%%%%%%%%%%%%%%%%%%%%%

\section{Introduction}

\label{sec:Intro}
%%%%%%%%%%%%%%%%%%%%%%%%%%%%%%%%%%%%%%%%%%%%%%%%%%%%%%%%%%%

Experimental discoveries of new four $X$ structures reported by the LHCb,
ATLAS and CMS Collaborations had important consequences for physics of
exotic four-quark mesons composed of exclusively heavy quarks \cite%
{LHCb:2020bwg,Bouhova-Thacker:2022vnt,CMS:2023owd}. These $X$ states have
the masses $6.2-7.3\ \mathrm{GeV}$ and are presumably scalar resonances with
the content $cc\overline{c}\overline{c}$. Note that their interpretations as
kinematical effects were suggested as well.

Exotic mesons composed of four heavy quarks were studied during long time as
interesting structures. The reason is that neither the quark-parton model
nor quantum chromodynamics (QCD) forbid existence of states beyond the $q%
\overline{q}$ and $qq^{\prime }q^{\prime \prime }$ spectroscopy.
Experimental observations of recent years supported interest to such
unconventional hadrons.

The tetraquarks $bc\overline{b}\overline{c}$ with various quantum numbers
are such particles. It is possible that they will be soon observed in
experiments \cite{Carvalho:2015nqf,Abreu:2023wwg}. These tetraquarks may
have different internal organization. First, the colored diquark $bc$ and
antidiquark $\overline{b}\overline{c}$ may form a colorless
diquark-antidiquark state with various quantum numbers. Second, $\overline{c}%
b$ and $b\overline{c}$ establish conventional $B_{c}$ mesons, and a
tetraquark emerges as the hadronic $B_{c}B_{c}$ molecule state.

The diquark-antidiquark structures $bc\overline{b}\overline{c}$ with
numerous quantum numbers were intensively investigated in the literature
\cite{Faustov:2022mvs,Wu:2016vtq,Liu:2019zuc,Chen:2019vrj,Bedolla:2019zwg,
Cordillo:2020sgc,Weng:2020jao,Yang:2021zrc,Hoffer:2024alv}. In these
articles the authors concentrated mainly on their masses and computed them
 by employing different methods. It should be noted that the decays
of the tetraquarks $bc\overline{b}\overline{c}$, as usual, remained beyond
the scope of these studies. In our publications \cite%
{Agaev:2024wvp,Agaev:2024mng,Agaev:2024qbh} we considered the
diquark-antidiquark states $bc\overline{b}\overline{c}$ with spin-parities $%
J^{\mathrm{P}}=0^{+}$, $1^{+}$, $2^{+}$ and calculated their masses and
widths.

The molecular structures $B_{c}B_{c}$ were objects of explorations as well
\cite{Liu:2023gla,Liu:2024pio,Agaev:2025wdj,Agaev:2025fwm}. In Refs.\ \cite%
{Liu:2023gla,Liu:2024pio} they were studied in the framework of the
coupled-channel unitary approach. The masses and widths of the scalar $%
B_{c}^{+}B_{c}^{-}$ and axial-vector $(B_{c}^{\ast
+}B_{c}^{-}+B_{c}^{+}B_{c}^{\ast -})/2$ molecules were calculated in our
works as well \cite{Agaev:2025wdj,Agaev:2025fwm}. To find parameters of
these molecular states, we applied QCD sum rule (SR) method \cite%
{Shifman:1978bx,Shifman:1978by}. Our predictions for the masses $(12.725\pm
0.085)~\mathrm{GeV}$ and $(12.77\pm 0.06)~\mathrm{GeV}$ of these compounds
demonstrated that they are unstable against strong decays into ordinary
mesons. Their full decay widths $(155\pm 23)~\mathrm{MeV}$ and $(93\pm 14)~%
\mathrm{MeV}$, respectively, characterize them as relatively broad exotic
mesons.

In this article, we continue our studies of $B_{c}$ mesons' molecules and
explore the hadronic tensor state $\mathcal{M}_{\mathrm{T}}=B_{c}^{\ast
+}B_{c}^{\ast -}$. We are going to calculate its mass $m$ and current
coupling $\Lambda $ (pole residue) using the two-point sum rule method. The
three-point sum rule approach is applied to treat numerous decay channels of
this state. Technical tools of this approach are required to evaluate the
form factors that describe strong interaction of particles at $\mathcal{M}_{%
\mathrm{T}}M_{1}M_{2}$ vertices, where $M_{1}$ and $M_{2}$ are final-state
mesons with appropriate charges and spin-parities. In its turns, these form
factors at the mass shell of one of the mesons allow us to estimate the
strong couplings at $\mathcal{M}_{\mathrm{T}}M_{1}M_{2}$.

Note that there are different mechanisms that trigger decay of the molecule $%
\mathcal{M}_{\mathrm{T}}$. In fact, $\mathcal{M}_{\mathrm{T}}$ can decay to
heavy mesons $J/\psi \Upsilon $ and $\eta _{b}\eta _{c}$, $B_{c}^{\ast
+}B_{c}^{\ast -}$ and $B_{c}^{+}B_{c}^{-}$, in which all constituent quarks
of the hadronic molecule participate in forming of the final-state
particles. These processes are dominant decay modes of $\mathcal{M}_{\mathrm{%
T}}$. Another type of decays are due to subprocesses $b\overline{b}$ or $c%
\overline{c}\rightarrow $ $q\overline{q}$, $s\overline{s}$ inside of $%
\mathcal{M}_{\mathrm{T}}$ with subsequent creation of $D_{(s)}D_{(s)}$ and $%
B_{(s)}B_{(s)}$ pairs provided the mass $m$ and quantum numbers of $D_{(s)}$
and $B_{(s)}$ mesons make such processes kinematically possible \cite%
{Becchi:2020mjz,Becchi:2020uvq,Agaev:2023ara}.

The current paper is divided into seven sections: In Sec.\ \ref{sec:Mass},
we compute the spectroscopic parameters of the hadronic tensor molecule $%
\mathcal{M}_{\mathrm{T}}$. The decays $\mathcal{M}_{\mathrm{T}}\rightarrow
J/\psi \Upsilon $ and $\eta _{b}\eta _{c}$ are considered in Sec.\ \ref%
{sec:Widths1}, in which we evaluate their partial widths. The decays into $%
B_{c}^{\ast +}B_{c}^{\ast -}$ and $B_{c}^{+}B_{c}^{-}$ mesons are explored
in Sec. \ref{sec:Widths2}. The widths of the processes $\mathcal{M}_{\mathrm{%
T}}\rightarrow D^{(\ast )+}D^{(\ast )-}$, $D^{(\ast )0}\overline{D}^{(\ast
)0}$ and $D_{s}^{(\ast )+}D_{s}^{(\ast )-}$ are examined in Sec.\ \ref%
{sec:Widths3}. Section \ref{sec:Widths4} is devoted to analysis of $\mathcal{%
M}_{\mathrm{T}}$ molecule's decays to $B$ meson pairs, i.e. to decays $%
B^{(\ast )+}B^{(\ast )-}$, $B^{(\ast )0}\overline{B}^{(\ast )0}$, and $%
B_{s}^{0}\overline{B}_{s}^{0}.$ Here, we also evaluate the full width of the
molecule $\mathcal{M}_{\mathrm{T}}$. Discussion of obtained results and
related conclusions are made in Sec.\ \ref{sec:Conc}.

%%%%%%%%%%%%%%%%%%%%%%%%%%%%%%%%%%%%%%%%%%%%%%%%%%%%%%%%%%%%%%%%%

\section{Spectroscopic parameters $m$ and $\Lambda $ of the molecule $%
\mathcal{M}_{\mathrm{T}}$}

\label{sec:Mass}
%%%%%%%%%%%%%%%%%%%%%%%%%%%%%%%%%%%%%%%%%%%%%%%%%%%%%%%%%%%

The mass $m$ and current coupling (pole residue) $\Lambda $ of the hadronic
molecule $\mathcal{M}_{\mathrm{T}}$ are important quantities that determine
properties of this state. One of the powerful nonperturbative approaches to
determine them is the two-point sum rule method. In this context we have to
derive the SRs for these parameters. For these purposes, we analyze the
correlation function
\begin{equation}
\Pi _{\mu \nu \alpha \beta }(p)=i\int d^{4}xe^{ipx}\langle 0|\mathcal{T}%
\{J_{\mu \nu }(x)J_{\alpha \beta }^{\dag }(0)\}|0\rangle ,  \label{eq:CF1}
\end{equation}%
where $J_{\mu \nu }(x)$ is the interpolating current for the tensor molecule
$\mathcal{M}_{\mathrm{T}}$. The time-ordered product of two currents in Eq.\
(\ref{eq:CF1}) is denoted by $\mathcal{T}$.

The current $J_{\mu \nu }(x)$ for the molecule $\mathcal{M}_{\mathrm{T}%
}=B_{c}^{\ast +}B_{c}^{\ast -}$ has the form
\begin{equation}
J_{\mu \nu }(x)=[\overline{b}_{a}(x)\gamma _{\mu }c_{a}(x)][\overline{c}%
_{b}(x)\gamma _{\nu }b_{b}(x)],  \label{eq:CR1}
\end{equation}%
where $a$ and $b$ are the color indices. The current $J_{\mu \nu }$
corresponds to the state with quantum numbers $J^{\mathrm{PC}}=2^{++}$.

The SRs for $m$ and $\Lambda $ can be obtained by equating $\Pi _{\mu \nu
\alpha \beta }^{\mathrm{Phys}}(p)$ and $\Pi _{\mu \nu \alpha \beta }^{%
\mathrm{OPE}}(p)$. The first function $\Pi _{\mu \nu \alpha \beta }^{\mathrm{%
Phys}}(p)$ is the correlator expressed using the physical parameters of the
molecule $\mathcal{M}_{\mathrm{T}}$. To find it, one inserts into Eq.\ (\ref%
{eq:CF1}) a full set of states with the quark content and quantum numbers of
$\mathcal{M}_{\mathrm{T}}$, and carries out integration over $x$. Finally, $%
\Pi _{\mu \nu \alpha \beta }^{\mathrm{Phys}}(p)$ becomes equal to
\begin{eqnarray}
\Pi _{\mu \nu \alpha \beta }^{\mathrm{Phys}}(p) &=&\frac{\langle 0|J_{\mu
\nu }|\mathcal{M}_{\mathrm{T}}(p,\epsilon )\rangle \langle \mathcal{M}_{%
\mathrm{T}}(p,\epsilon )|J_{\alpha \beta }^{\dag }|0\rangle }{m^{2}-p^{2}}
\notag \\
&&+\cdots .  \label{eq:Phys1}
\end{eqnarray}%
Here, the term shown in Eq.\ (\ref{eq:Phys1}) is due to the ground-level
molecule $\mathcal{M}_{\mathrm{T}}$, while the ellipses denote contributions
of higher resonances and continuum states. To proceed with computations, we
use the matrix element
\begin{equation}
\langle 0|J_{\mu \nu }|\mathcal{M}_{\mathrm{T}}(p,\epsilon (p)\rangle
=\Lambda \epsilon _{\mu \nu }^{(\lambda )}(p),  \label{eq:ME1}
\end{equation}%
where $\epsilon =\epsilon _{\mu \nu }^{(\lambda )}(p)$ is the polarization
tensor of the molecule $\mathcal{M}_{\mathrm{T}}$. We replace in $\Pi _{\mu
\nu \alpha \beta }^{\mathrm{Phys}}(p)$ matrix elements in the correlator
Eq.\ (\ref{eq:Phys1}) by Eq. (\ref{eq:ME1}) and perform required operations.
Our computations yield
\begin{eqnarray}
\Pi _{\mu \nu \alpha \beta }^{\mathrm{Phys}}(p) &=&\frac{\Lambda ^{2}}{%
m^{2}-p^{2}}\left\{ \frac{1}{2}\left( g_{\mu \alpha }g_{\nu \beta }+g_{\mu
\beta }g_{\nu \alpha }\right) \right.  \notag \\
&&\left. +\text{ other terms}\right\} +\cdots ,  \label{eq:Phys2}
\end{eqnarray}%
where dots stand for higher resonances and continuum states. Note that there
are various Lorentz structures in Eq.\ (\ref{eq:Phys2}). The component $\sim
(g_{\mu \alpha }g_{\nu \beta }+g_{\mu \beta }g_{\nu \alpha })$ is formed due
to only the spin-$2$ particle. Therefore, we choose for our studies this
component of the correlation function and denote by $\Pi ^{\mathrm{Phys}%
}(p^{2})$ the related invariant amplitude.

To compute $\Pi _{\mu \nu \alpha \beta }^{\mathrm{OPE}}(p)$ one inserts $%
J_{\mu \nu }(x)$ into Eq.\ (\ref{eq:CF1}) and contract quark fields. In the
case under discussion, we get
\begin{eqnarray}
&&\Pi _{\mu \nu \alpha \beta }^{\mathrm{OPE}}(p)=i\int d^{4}xe^{ipx}\mathrm{%
Tr}\left[ \gamma _{\mu }S_{c}^{aa^{\prime }}(x)\gamma _{\beta
}S_{b}^{a^{\prime }a}(-x)\right]  \notag \\
&&\times \mathrm{Tr}\left[ \gamma _{\nu }S_{b}^{bb^{\prime }}(x)\gamma
_{\alpha }S_{c}^{b^{\prime }b}(-x)\right] ,  \label{eq:QCD1}
\end{eqnarray}%
where $S_{b(c)}(x)$ are heavy-quark propagators \cite{Agaev:2020zad}. The
correlator $\Pi _{\mu \nu \alpha \beta }^{\mathrm{OPE}}(p)$ obtained by this
way has to be calculated in the operator product expansion ($\mathrm{OPE}$)
with certain accuracy.

Having extracted the structure $(g_{\mu \alpha }g_{\nu \beta }+g_{\mu \beta
}g_{\nu \alpha })$ from $\Pi _{\mu \nu \alpha \beta }^{\mathrm{OPE}}(p)$ and
labeled the relevant invariant amplitude by $\Pi ^{\mathrm{OPE}}(p^{2})$,
one finds SRs for $m$ and $\Lambda $, which reads
\begin{equation}
m^{2}=\frac{\Pi ^{\prime }(M^{2},s_{0})}{\Pi (M^{2},s_{0})},  \label{eq:Mass}
\end{equation}%
and
\begin{equation}
\Lambda ^{2}=e^{m^{2}/M^{2}}\Pi (M^{2},s_{0}).  \label{eq:Coupl}
\end{equation}%
The function $\Pi (M^{2},s_{0})$ is the amplitude $\Pi ^{\mathrm{OPE}%
}(p^{2}) $ obtained after the Borel transformation and continuum
subtraction, whereas $\Pi ^{\prime }(M^{2},s_{0})$ is its derivative over $%
d/d(-1/M^{2})$. The Borel transformation is applied to suppress contribution
of higher resonances and continuum states, whereas these terms are
subtracted from the $\Pi (M^{2},\infty )$ using ideas of the quark-hadron
duality and continuum threshold parameter $s_{0}$. As a result, $\Pi
(M^{2},s_{0})$ depends on $s_{0}$ and on the Borel parameter $M^{2}$. It has
the following form.
\begin{equation}
\Pi (M^{2},s_{0})=\int_{4(m_{b}+m_{c})^{2}}^{s_{0}}ds\rho ^{\mathrm{OPE}%
}(s)e^{-s/M^{2}}+\Pi (M^{2}).  \label{eq:CorrF}
\end{equation}%
The spectral density $\rho ^{\mathrm{OPE}}(s)$ is calculated as the
imaginary piece of the amplitude $\Pi ^{\mathrm{OPE}}(p^{2})$. The
nonperturbative contribution $\Pi (M^{2})$ is found straightly from $\Pi ^{%
\mathrm{OPE}}(p)$ and encompasses terms do not included into $\rho ^{\mathrm{%
OPE}}(s)$.

To conduct numerical calculations one needs to choose input parameters in
Eqs.\ (\ref{eq:Mass}) and (\ref{eq:Coupl}). The masses $m_{c}$ and $m_{b}$
of the quarks as well as the condensate $\langle \alpha _{s}G^{2}/\pi
\rangle $ are universal quantities. In the current article, we employ
\begin{eqnarray}
&&m_{c}=(1.2730\pm 0.0046)~\mathrm{GeV},  \notag \\
&&m_{b}=(4.183\pm 0.007)~\mathrm{GeV},  \notag \\
&&\langle \alpha _{s}G^{2}/\pi \rangle =(0.012\pm 0.004)~\mathrm{GeV}^{4}.
\label{eq:GluonCond}
\end{eqnarray}%
The quark masses $m_{c}$ and $m_{b}$ are calculated in the $\overline{%
\mathrm{MS}}$ scheme \cite{PDG:2024}. The condensate $\langle \alpha
_{s}G^{2}/\pi \rangle $ was estimated in Refs.\ \cite%
{Shifman:1978bx,Shifman:1978by} from studies of different processes.

The Borel $M^{2}$ and continuum subtraction $s_{0}$ parameters depend on the
process under analysis and should satisfy the usual restrictions of the sum
rule calculations. Prevalence of the pole contribution ($\mathrm{PC}$) in
the mass $m$ and current coupling $\Lambda $, their independence on $M^{2}$
and $s_{0}$, as well as convergence of $\mathrm{OPE}$ are decisive for the
credible SR analysis. To obey these constraints, we impose on $M^{2}$ and $%
s_{0}$ the following requirements. First, the pole contribution
\begin{equation}
\mathrm{PC}=\frac{\Pi (M^{2},s_{0})}{\Pi (M^{2},\infty )},  \label{eq:PC}
\end{equation}%
has to satisfy $\mathrm{PC}>0.5$ which leads to its dominance in $m$ and $%
\Lambda $. The correlation function contains only the dimension-$4$ term $%
\Pi ^{\mathrm{Dim4}}(M^{2},s_{0})$, therefore we demand fulfilment of the
constraint $|\Pi ^{\mathrm{Dim4}}(M^{2},s_{0})|\leq 0.05|\Pi (M^{2},s_{0})|$%
: This ensures the convergence of $\mathrm{OPE}$. The maximum of the Borel
parameter is fixed by Eq.\ (\ref{eq:PC}), whereas convergence of $\mathrm{OPE%
}$ permits us to obtain its minimal value.

\begin{figure}[h]
\includegraphics[width=8.5cm]{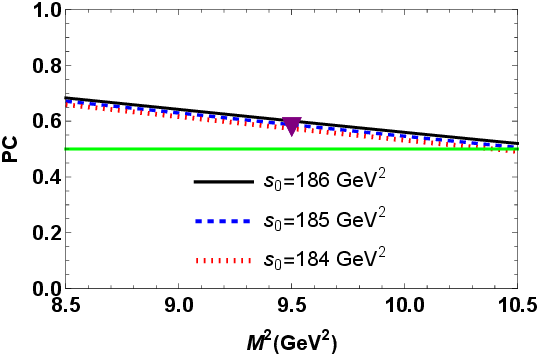}
\caption{The $\mathrm{PC}$ as a function of $M^{2}$ at various $s_{0}$. The
border $\mathrm{PC}=0.5$ is plotted by the horizontal line. The red triangle
fixes the point $M^{2}=9.5~\mathrm{GeV}^{2}$ and $s_{0}=185~\mathrm{GeV}^{2}$%
. }
\label{fig:PC}
\end{figure}

\begin{widetext}

\begin{figure}[htbp]
\begin{center}
\includegraphics[totalheight=6cm,width=8cm]{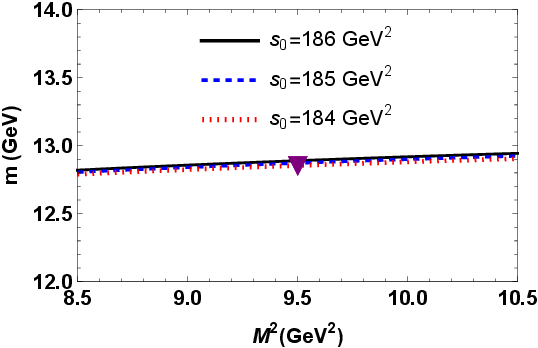}
\includegraphics[totalheight=6cm,width=8cm]{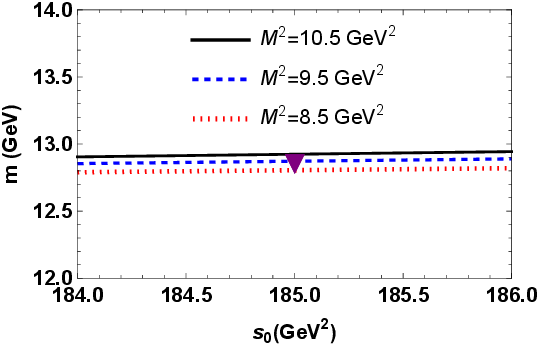}
\end{center}
\caption{Mass $m$ as a function of the Borel  $M^{2}$ (left panel), and continuum threshold $s_0$ parameters (right panel).}
\label{fig:Mass}
\end{figure}

\end{widetext}

Numerical computations are carried out for different values of $M^{2}$ and $%
s_{0}$. Examination of obtained results permits us to determine regions for $%
M^{2}$ and $s_{0}$, where all restrictions are satisfied. We find that the
regions
\begin{equation}
M^{2}\in \lbrack 8.5,10.5]~\mathrm{GeV}^{2},\ s_{0}\in \lbrack 184,186]~%
\mathrm{GeV}^{2}  \label{eq:Wind1}
\end{equation}%
comply with these constraints. In fact, at maximal and minimal values of $%
M^{2}$ the pole contribution on the average in $s_{0}$ is $\mathrm{PC}%
\approx 0.50$ and $\mathrm{PC}\approx 0.68$, correspondingly. The
nonperturbative term is negative and establishes at $M^{2}=9.5~\mathrm{GeV}%
^{2}$ approximately $1\%$ of the full result. The parameter $\mathrm{PC}$ as
a function of $M^{2}$ is depicted in Fig.\ \ref{fig:PC}, where all curves
overshoot the border $\mathrm{PC}=0.5$.

We compute mean values of $m$ and $\Lambda $ over the regions Eq.\ (\ref%
{eq:Wind1}) and find
\begin{eqnarray}
&&m=(12.87\pm 0.08)~\mathrm{GeV},  \notag \\
&&\Lambda =(6.70\pm 0.78)\times 10^{-2}~\mathrm{GeV}^{5}.  \label{eq:Result1}
\end{eqnarray}%
The results in Eq.\ (\ref{eq:Result1}) amount to SR predictions at the point
$M^{2}=9.5~\mathrm{GeV}^{2}$ and $s_{0}=185~\mathrm{GeV}^{2}$, where $%
\mathrm{PC}\approx 0.59$. This guaranties prevalence of $\mathrm{PC}$ in the
quantities $m$ and $\Lambda $. The main sources of ambiguities in Eq.\ (\ref%
{eq:Result1}) are parameters $M^{2}$ and $s_{0}$: Errors generated by
uncertainties in the quark masses and gluon condensate are negligibly small.
Relevant errors constitutes $\pm 0.6\%$ of the mass $m$, which proves
reliability of the extracted result. Such accuracy is achieved due to the
sum rule for $m$, Eq.\ (\ref{eq:Mass}), which is a ratio of correlators.
Consequently, variations of these functions due to changes of $M^{2}$, and $%
s_{0}$ recompense in $m$ and stabilize the result. Ambiguities of the
current coupling $\Lambda $ are equal to $\pm 11.6\%$ remaining nevertheless
inside borders allowable for SR studies. In Fig.\ \ref{fig:Mass}, we depict $%
m$ as a function of $M^{2}$ and $s_{0}$.

%%%%%%%%%%%%%%%%%%%%%%%%%%%%%%%%%%%%%%%%%%%%%%%%%%%%%%%%%%%%%%%%%

\section{Processes $\mathcal{M}_{\mathrm{T}}\rightarrow J/\protect\psi %
\Upsilon $, $\protect\eta _{b}\protect\eta _{c}$}

\label{sec:Widths1}

%%%%%%%%%%%%%%%%%%%%%%%%%%%%%%%%%%%%%%%%%%%%%%%%%%%%%%%%%%%

Parameters of the hadronic molecule $\mathcal{M}_{\mathrm{T}}$ fix its
kinematically allowed decay modes. Decays to mesons $J/\psi \Upsilon $ and $%
\eta _{b}\eta _{c}$ are such channels. In fact, thresholds for these
processes are $12.558~\mathrm{GeV}$ and $12.383~\mathrm{GeV}$, respectively.
Here, we concentrate on the decays $\mathcal{M}_{\mathrm{T}}\rightarrow
J/\psi \Upsilon $ and $\mathcal{M}_{\mathrm{T}}\rightarrow \eta _{b}\eta
_{c} $.

%%%%%%%%%%%%%%%%%%%%%%%%%%%%%%%%%%%%%%%%%%%%%%%%%%%%%%%%%%%%%%%%%%%%%%%

\subsection{Decay $\mathcal{M}_{\mathrm{T}}\rightarrow J/\protect\psi %
\Upsilon $}

%%%%%%%%%%%%%%%%%%%%%%%%%%%%%%%%%%%%%%%%%%%%%%%%%%%%%%%%%%%%%%%%%%%%%%%

First, we explore the channel $\mathcal{M}_{\mathrm{T}}\rightarrow J/\psi
\Upsilon $ and compute its partial width. For these purposes, we should
determine the strong coupling $g_{1}$ at the vertex $\mathcal{M}_{\mathrm{T}%
}J/\psi \Upsilon $. This can be done by evaluating the form factor $%
g_{1}(q^{2})$ at the mass shell $q^{2}=m_{J/\psi }^{2}$.

We extract SR for the form factor $g_{1}(q^{2})$ from analysis of the
three-point correlator
\begin{eqnarray}
\Pi _{\mu \nu \alpha \beta }(p,p^{\prime }) &=&i^{2}\int
d^{4}xd^{4}ye^{ip^{\prime }y}e^{-ipx}\langle 0|\mathcal{T}\{J_{\mu
}^{\Upsilon }(y)  \notag \\
&&\times J_{\nu }^{J/\psi }(0)J_{\alpha \beta }^{\dagger }(x)\}|0\rangle ,
\label{eq:CF1a}
\end{eqnarray}%
with $J_{\mu }^{\Upsilon }(x)$ and $J_{\nu }^{J/\psi }(x)$ being the
interpolating currents of the vector quarkonia $\Upsilon $ and $J/\psi $,
respectively. They are given by expressions%
\begin{equation}
J_{\mu }^{\Upsilon }(x)=\overline{b}_{i}(x)\gamma _{\mu }b_{i}(x),\ J_{\nu
}^{J/\psi }(x)=\overline{c}_{j}(x)\gamma _{\nu }c_{j}(x),
\end{equation}%
where $i$ and $j$ are the color indices.

The physical side of the sum rule $\Pi _{\mu \nu \alpha \beta }^{\mathrm{Phys%
}}(p,p^{\prime })$ is found by rewriting Eq.\ (\ref{eq:CF1a}) using
parameters of the particles $\mathcal{M}_{\mathrm{T}}$, $J/\psi $, and $%
\Upsilon $. By considering only the ground-level states, we convert $\Pi
_{\mu \nu \alpha \beta }(p,p^{\prime })$ into the form
\begin{eqnarray}
&&\Pi _{\mu \nu \alpha \beta }^{\mathrm{Phys}}(p,p^{\prime })=\frac{\langle
0|J_{\mu }^{\Upsilon }|\Upsilon (p^{\prime },\varepsilon _{1})\rangle }{%
p^{\prime 2}-m_{\Upsilon }^{2}}\frac{\langle 0|J_{\nu }^{J/\psi }|J/\psi
(q,\varepsilon _{2})\rangle }{q^{2}-m_{J/\psi }^{2}}  \notag \\
&&\times \langle \Upsilon (p^{\prime },\varepsilon _{1})J/\psi
(q,\varepsilon _{2})|\mathcal{M}_{\mathrm{T}}(p,\epsilon )\rangle \frac{%
\langle \mathcal{M}_{\mathrm{T}}(p,\varepsilon )|J_{\alpha \beta }^{\dagger
}|0\rangle }{p^{2}-m^{2}}  \notag \\
&&+\cdots ,  \label{eq:TP1}
\end{eqnarray}%
where $m_{\Upsilon }=(9460.40\pm 0.10)~\mathrm{MeV}$ and $m_{J/\psi
}=(3096.900\pm 0.006)~\mathrm{MeV}$ are masses of the $\Upsilon $ and $%
J/\psi $ mesons \cite{PDG:2024}. Above, $\varepsilon _{1}$ and $\varepsilon
_{2}$ are the polarization vectors of the mesons $\Upsilon $ and $J/\psi $,
respectively.

Equation\ (\ref{eq:TP1}) can be written in a more simple form. To this end,
we employ the matrix elements%
\begin{eqnarray}
\langle 0|J_{\mu }^{\Upsilon }|\Upsilon (p^{\prime },\varepsilon
_{1})\rangle &=&f_{\Upsilon }m_{\Upsilon }\varepsilon _{1\mu }(p^{\prime }),
\notag \\
\langle 0|J_{\nu }^{J/\psi }|J/\psi (q,\varepsilon _{2})\rangle &=&f_{J/\psi
}m_{J/\psi }\varepsilon _{2\nu }(q).  \label{eq:C2}
\end{eqnarray}%
Here, $f_{\Upsilon }=(708\pm 8)~\mathrm{MeV}$ and $f_{J/\psi }=(411\pm 7)~%
\mathrm{MeV}$ are the experimental values of the mesons' decay constants\
\cite{Lakhina:2006vg}.

We also have to specify the matrix element $\langle \Upsilon (p^{\prime
},\varepsilon _{1})J/\psi (q,\varepsilon _{2})|\mathcal{M}_{\mathrm{T}%
}(p,\epsilon )\rangle $ which can be done by expressing it in terms of
parameters of the particles $\mathcal{M}_{\mathrm{T}}$, $\Upsilon $ and $%
J/\psi $ and corresponding form factors. Further analysis demonstrates that
the tensor-vector-vector vertex has the form \cite{Agaev:2024pil}
\begin{eqnarray}
&&\langle \Upsilon (p^{\prime },\varepsilon _{1})J/\psi (q,\varepsilon _{2})|%
\mathcal{M}_{\mathrm{T}}(p,\epsilon )\rangle =g_{1}(q^{2})\epsilon _{\tau
\rho }^{(\lambda )}\left[ (\varepsilon _{1}^{\ast }\cdot q)\right.  \notag \\
&&\left. \times \varepsilon _{2}^{\tau \ast }p^{\prime \rho }+(\varepsilon
_{2}^{\ast }\cdot p^{\prime })\varepsilon _{1}^{\ast \tau }q^{\rho
}-(p^{\prime }\cdot q)\varepsilon _{1}^{\tau \ast }\varepsilon _{2}^{\rho
\ast }-(\varepsilon _{1}^{\ast }\cdot \varepsilon _{2}^{\ast })p^{\prime
\tau }q^{\rho }\right] .  \notag \\
&&  \label{eq:TVV}
\end{eqnarray}%
Then, for $\Pi _{\mu \nu \alpha \beta }^{\mathrm{Phys}}(p,p^{\prime })$ we
get
\begin{eqnarray}
&&\Pi _{\mu \nu \alpha \beta }^{\mathrm{Phys}}(p,p^{\prime })=g_{1}(q^{2})%
\frac{\Lambda f_{\Upsilon }m_{\Upsilon }f_{J/\psi }m_{J/\psi }}{\left(
p^{2}-m^{2}\right) (p^{\prime 2}-m_{\Upsilon }^{2})(q^{2}-m_{J/\psi }^{2})}
\notag \\
&&\times \left[ p_{\beta }^{\prime }p_{\alpha }^{\prime }g_{\mu \nu }+\frac{1%
}{2}p_{\mu }p_{\alpha }^{\prime }g_{\beta \nu }+\frac{1}{2m^{2}}p_{\beta
}p_{\nu }p_{\mu }^{\prime }p_{\alpha }^{\prime }\right.  \notag \\
&&\left. +\text{ other structures}\right] +\cdots .
\end{eqnarray}

The correlation function $\Pi _{\mu \nu \alpha \beta }^{\mathrm{OPE}%
}(p,p^{\prime })$ is given by the formula
\begin{eqnarray}
&&\Pi _{\mu \nu \alpha \beta }^{\mathrm{OPE}}(p,p^{\prime })=\int
d^{4}xd^{4}ye^{ip^{\prime }y}e^{-ipx}\mathrm{Tr}\left[ \gamma _{\mu
}S_{b}^{ib}(y-x)\right.  \notag \\
&&\left. \times \gamma _{\alpha }S_{c}^{bj}(x)\gamma _{\nu
}S_{c}^{ja}(-x)\gamma _{\beta }S_{b}^{ai}(x-y)\right] .  \label{eq:CF3}
\end{eqnarray}%
We use the amplitudes $\Pi _{1}^{\mathrm{Phys}}(p^{2},p^{\prime 2},q^{2})$
and $\Pi _{1}^{\mathrm{OPE}}(p^{2},p^{\prime 2},q^{2})$ that correspond to
terms $\sim p_{\mu }p_{\alpha }^{\prime }g_{\beta \nu }$ in these
correlators, and determine SR for $g_{1}(q^{2})$. After standard
manipulations, we get%
\begin{equation}
g_{1}(q^{2})=\frac{2(q^{2}-m_{J/\psi }^{2})}{\Lambda f_{\Upsilon
}m_{\Upsilon }f_{J/\psi }m_{J/\psi }}e^{m^{2}/M_{1}^{2}}e^{m_{\Upsilon
}^{2}/M_{2}^{2}}\Pi _{1}(\mathbf{M}^{2},\mathbf{s}_{0},q^{2}).
\label{eq:SRG}
\end{equation}%
In Eq.\ (\ref{eq:SRG}), $\Pi _{1}(\mathbf{M}^{2},\mathbf{s}_{0},q^{2})$ is
the function $\Pi _{1}^{\mathrm{OPE}}(p^{2},p^{\prime 2},q^{2})$ after the
Borel transformations and continuum subtractions. As is clear it contains $%
\mathbf{M}^{2}=(M_{1}^{2},M_{2}^{2})$ and $\mathbf{s}_{0}=(s_{0},s_{0}^{%
\prime })$ where the pairs $(M_{1}^{2},s_{0})$ and $(M_{2}^{2},s_{0}^{\prime
})$ correspond to the molecule and $\Upsilon $ channels, respectively. It is
given by the formula%
\begin{eqnarray}
&&\Pi _{1}(\mathbf{M}^{2},\mathbf{s}_{0},q^{2})=\int_{4(m_b+m_c)
^{2}}^{s_{0}}ds\int_{4m_{b}^{2}}^{s_{0}^{\prime }}ds^{\prime }\rho
_{1}(s,s^{\prime },q^{2})  \notag \\
&&\times e^{-s/M_{1}^{2}-s^{\prime }/M_{2}^{2}}.  \label{eq:CorrF1}
\end{eqnarray}

Constraints used to fix the quantities $\mathbf{M}^{2}$ and $\mathbf{s}_{0}$
are standard for all sum rule investigations and have been outlined in Sec.\ %
\ref{sec:Mass}. Our calculations show that the regions in Eq.\ (\ref%
{eq:Wind1}) for the parameters $(M_{1}^{2},s_{0})$ and
\begin{equation}
M_{2}^{2}\in \lbrack 10,12]~\mathrm{GeV}^{2},\ s_{0}^{\prime }\in \lbrack
98,100]~\mathrm{GeV}^{2}.  \label{eq:Wind3}
\end{equation}%
for $(M_{2}^{2},s_{0}^{\prime })$ satisfy these requirements. Let us note
that $s_{0}^{\prime }$ is bounded by the mass $m_{\Upsilon (2S)}=(10023.4\pm
0.5)~\mathrm{MeV}\ $of the excited state $\Upsilon (2S)$, i.e., $%
s_{0}^{\prime }<m_{\Upsilon (2S)}^{2}$.

\begin{figure}[h]
\includegraphics[width=8.5cm]{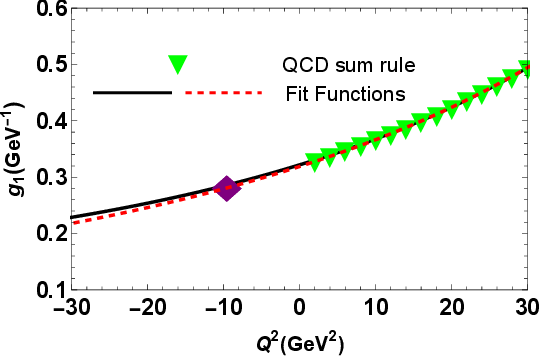}
\caption{QCD data and fit functions for $g_{1}(Q^{2})$ (solid line) and $%
\overline{g}_{1}(Q^{2})$ (dashed line). The diamond fixes the point $%
Q^{2}=-m_{J/\protect\psi }^{2}$ where $g_{1}$ and $\overline{g}_{1}$ have
been estimated. }
\label{fig:Fit}
\end{figure}

The sum rule gives reliable predictions for the form factor $g_{1}(q^{2})$
in the Euclidean region $q^{2}<0$. But $g_{1}(q^{2})$ determines the
coupling $g_{1}$ at the mass shell $q^{2}=m_{J/\psi }^{2}$. Therefore, it is
convenient to employ the function $g_{1}(Q^{2})$ with $Q^{2}=-q^{2}$ and use
it in our analysis. The SR results for $g_{1}(Q^{2})$ are plotted in Fig.\ %
\ref{fig:Fit}, where $Q^{2}$ varies in the region $Q^{2}=2-30~\mathrm{GeV}%
^{2}$.

Above it has been noted that $g_{1}$ should be extracted at $q^{2}=m_{J/\psi
}^{2}$, i.e., at $Q^{2}=-m_{J/\psi }^{2}$ where the SR method can not be
applied. Therefore, we use the function $\mathcal{G}_{1}(Q^{2})$ which at $%
Q^{2}>0$ is equal to the SR data $g_{1}(Q^{2})$, but can be extrapolated to
the region $Q^{2}<0$. To this end, we introduce the function
\begin{equation}
\mathcal{G}_{i}(Q^{2})=\mathcal{G}_{i}^{0}\mathrm{\exp }\left[ c_{i}^{1}%
\frac{Q^{2}}{m^{2}}+c_{i}^{2}\left( \frac{Q^{2}}{m^{2}}\right) ^{2}\right] ,
\label{eq:FitF}
\end{equation}%
where $\mathcal{G}_{i}^{0}$, $c_{i}^{1}$, and $c_{i}^{2}$ are fitted
constants. From confronting of the SR data and Eq.\ (\ref{eq:FitF}), it is
not difficult to find
\begin{equation}
\mathcal{G}_{1}^{0}=0.32~\mathrm{GeV}^{-1},c_{1}^{1}=2.13,\text{ }%
c_{1}^{2}=1.34.  \label{eq:FF1}
\end{equation}%
The function $\mathcal{G}_{1}(Q^{2})$ is plotted in Fig.\ \ref{fig:Fit},
where its agreement with the QCD data is clear. For the strong coupling $%
g_{1}$, we find
\begin{equation}
g_{1}\equiv \mathcal{G}_{1}(-m_{J/\psi }^{2})=(2.86\pm 0.50)\times 10^{-1}\
\mathrm{GeV}^{-1}.  \label{eq:g1}
\end{equation}

Partial width of the decay $\mathcal{M}_{\mathrm{T}}\rightarrow J/\psi
\Upsilon $ is given by the formula%
\begin{equation}
\Gamma \left[ \mathcal{M}_{\mathrm{T}}\rightarrow J/\psi \Upsilon \right]
=g_{1}^{2}\frac{\lambda _{1}}{40\pi m^{2}}|M_{1}|^{2},  \label{eq:PDw2}
\end{equation}%
where%
\begin{eqnarray}
&&|M_{1}|^{2}=\frac{1}{6m^{4}}\left[ m_{J/\psi }^{8}+m_{J/\psi
}^{6}(m^{2}-4m_{\Upsilon }^{2})+(m^{2}-m_{\Upsilon }^{2})^{2}\right.  \notag
\\
&&\times (6m^{4}+3m^{2}m_{\Upsilon }^{2}+m_{\Upsilon }^{4})+m_{J/\psi
}^{4}(m^{4}-m^{2}m_{\Upsilon }^{2}+6m_{\Upsilon }^{4})  \notag \\
&&\left. -m_{J/\psi }^{2}(9m^{6}-34m^{4}m_{\Upsilon }^{2}+m^{2}m_{\Upsilon
}^{4}+4m_{\Upsilon }^{6})\right] .  \label{eq:M1}
\end{eqnarray}%
In Eq.\ (\ref{eq:PDw2}), $\lambda _{1}=\lambda (m,m_{\Upsilon },m_{J/\psi })$
is determined by means of the expression
\begin{equation}
\lambda (x,y,z)=\frac{\sqrt{%
x^{4}+y^{4}+z^{4}-2(x^{2}y^{2}+x^{2}z^{2}+y^{2}z^{2})}}{2x}.
\end{equation}%
Finally, one gets
\begin{equation}
\Gamma \left[ \mathcal{M}_{\mathrm{T}}\rightarrow J/\psi \Upsilon \right]
=(23.7\pm 6.8)~\mathrm{MeV}.  \label{eq:DW2}
\end{equation}

The errors above originate from the ambiguities in the strong coupling $g_{1}
$ and those of the particles' masses in Eq.\ (\ref{eq:PDw2}). A potential
source of theoretical errors is also a choice of the interpolation function $%
\mathcal{G}_{i}(Q^{2})$ in the form of Eq.\ (\ref{eq:FitF}). It should be
emphasized that alternative forms for this function, in general, may lead to
deviations from Eq.\ (\ref{eq:g1}), but as usual they are very small and can
be neglected. In fact, let us consider the fit function in the following
analytical form%
\begin{equation}
\overline{\mathcal{G}}_{1}(Q^{2})=\frac{\overline{\mathcal{G}}_{i}^{0}}{%
\left( 1-\frac{Q^{2}}{m^{2}}\right) \left( 1-\sigma _{1}\frac{Q^{2}}{m^{2}}%
+\sigma _{2}\left( \frac{Q^{2}}{m^{2}}\right) ^{2}\right) },
\label{eq:FitF2}
\end{equation}%
where $\overline{\mathcal{G}}_{1}^{0}$, $\sigma _{1}$ and $\sigma _{2}$ are
fitting constants.

Then, using the QCD data and Eq. (\ref{eq:FitF2}), one finds the parameters $%
\overline{\mathcal{G}}_{1}^{0}=0.319$, $\sigma _{1}=1.244$ and $\sigma
_{2}=0.402$ in  $\overline{\mathcal{G}}_{1}(Q^{2})$. The function $\overline{%
\mathcal{G}}_{1}(Q^{2})$ is also depicted in Fig.\ \ref{fig:Fit}, where one
can see its nice agreement with SR data. Having used this function, we get $%
\overline{g}_{1}=0.281~\mathrm{GeV}^{-1}$. As is seen, a difference $g_{1}-%
\overline{g}_{1}=0.005$ is an order of magnitude smaller than ambiguities $%
\pm 0.05$ of $g_{1}$ from Eq.\ (\ref{eq:g1}).Therefore, throughout this work
we employ Eq.\ (\ref{eq:FitF}) and neglect small effects due to alternative
extrapolating functions.

%%%%%%%%%%%%%%%%%%%%%%%%%%%%%%%%%%%%%%%%%%%%%%%%%%%%%%%%%%%%%%%%%%5

\subsection{$\mathcal{M}_{\mathrm{T}}\rightarrow \protect\eta _{b}\protect%
\eta _{c}$}

%%%%%%%%%%%%%%%%%%%%%%%%%%%%%%%%%%%%%%%%%%%%%%%%%%%%%%%%%%%%%%%%%%%%%%%

Parameters of this decay depend on the strong coupling $g_{2}$ at the vertex
$\mathcal{M}_{\mathrm{T}}\eta _{b}\eta _{c}$. To find the relevant form
factor $g_{2}(q^{2})$, one has to consider the correlator%
\begin{eqnarray}
\Pi _{\mu \nu }(p,p^{\prime }) &=&i^{2}\int d^{4}xd^{4}ye^{ip^{\prime
}y}e^{-ipx}\langle 0|\mathcal{T}\{\ J^{\eta _{b}}(y)  \notag \\
&&\times J^{\eta _{c}}(0)J_{\mu \nu }^{\dagger }(x)\}|0\rangle .
\label{eq:CF7}
\end{eqnarray}%
The interpolating currents of the quarkonia $\eta _{c}$ and $\eta _{b}$ are
\begin{equation}
J^{\eta _{c}}(x)=\overline{c}_{i}(x)i\gamma _{5}c_{i}(x),\ J^{\eta _{b}}(x)=%
\overline{b}_{j}(x)i\gamma _{5}b_{j}(x).  \label{eq:C3}
\end{equation}

To calculate $\Pi _{\mu \nu }^{\mathrm{Phys}}(p,p^{\prime })$ it is
necessary to introduce the matrix elements
\begin{eqnarray}
&&\langle 0|J^{\eta _{b}}|\eta _{b}(p^{\prime })\rangle =\frac{f_{\eta
_{b}}m_{\eta _{b}}^{2}}{2m_{b}},  \notag \\
&&\langle 0|J^{\eta _{c}}|\eta _{c}(q)\rangle =\frac{f_{\eta _{c}}m_{\eta
_{c}}^{2}}{2m_{c}}.  \label{eq:ME4}
\end{eqnarray}%
Here, $f_{\eta _{b}}$, $m_{\eta _{b}}$ and $f_{\eta _{c}}$, $m_{\eta _{c}}$
are the decay constants and masses of the mesons $\eta _{b}$ and $\eta _{c}$%
, respectively. The vertex $\mathcal{M}_{\mathrm{T}}\eta _{b}\eta _{c}$ is
determined by the expression \cite{Agaev:2024pil}%
\begin{equation}
\langle \eta _{b}(p^{\prime })\eta _{c}(q)|\mathcal{M}_{\mathrm{T}%
}(p,\epsilon )\rangle =g_{2}(q^{2})\epsilon _{\alpha \beta }^{(\lambda
)}(p)p^{\prime \alpha }p^{\prime \beta }.
\end{equation}%
Then $\Pi _{\mu \nu }^{\mathrm{Phys}}(p,p^{\prime })$ is
\begin{eqnarray}
&&\Pi _{\mu \nu }^{\mathrm{Phys}}(p,p^{\prime })=g_{2}(q^{2})\frac{\Lambda
f_{\eta _{b}}m_{\eta _{b}}^{2}f_{\eta _{c}}m_{\eta _{c}}^{2}}{%
4m_{b}m_{c}\left( p^{2}-m^{2}\right) (p^{\prime 2}-m_{\eta _{b}}^{2})}
\notag \\
&&\times \frac{1}{(q^{2}-m_{\eta _{c}}^{2})}\left[ \frac{m^{4}-2m^{2}(m_{%
\eta _{b}}^{2}+q^{2})+(m_{\eta _{b}}^{2}-q^{2})^{2}}{12m^{2}}g_{\mu \nu
}\right.  \notag \\
&&\left. +p_{\mu }^{\prime }p_{\nu }^{\prime }+\text{other terms}\right] .
\label{eq:CF4}
\end{eqnarray}%
The function $\Pi _{\mu \nu }^{\mathrm{OPE}}(p,p^{\prime })$ reads
\begin{eqnarray}
&&\Pi _{\mu \nu }^{\mathrm{OPE}}(p,p^{\prime })=i\int
d^{4}xd^{4}ye^{ip^{\prime }y}e^{-ipx}\mathrm{Tr}\left[ \gamma
_{5}S_{b}^{ib}(y-x)\right.  \notag \\
&&\left. \times \gamma _{\mu }S_{c}^{bj}(x)\gamma _{5}S_{c}^{ja}(-x)\gamma
_{\nu }S_{b}^{ai}(x-y)\right] .  \label{eq:CF5}
\end{eqnarray}%
The terms in $\Pi _{\mu \nu }^{\mathrm{Phys}}(p,p^{\prime })$ and $\Pi _{\mu
\nu }^{\mathrm{OPE}}(p,p^{\prime })$ have the identical Lorentz
compositions. To derive the sum rule for $g_{2}(q^{2})$, we choose
components $\sim p_{\mu }^{\prime }p_{\nu }^{\prime }$ and utilize related
amplitudes $\Pi _{2}^{\mathrm{Phys}}(p^{2},p^{\prime 2},q^{2})$ and $\Pi
_{2}^{\mathrm{OPE}}(p^{2},p^{\prime 2},q^{2})$. As a result, we get

\begin{eqnarray}
&&g_{2}(q^{2})=\frac{4m_{c}m_{b}(q^{2}-m_{\eta _{c}}^{2})}{\Lambda f_{\eta
_{c}}m_{\eta _{c}}^{2}f_{\eta _{b}}m_{\eta _{b}}^{2}}%
e^{m^{2}/M_{1}^{2}}e^{m_{\eta _{b}}^{2}/M_{2}^{2}}  \notag \\
&&\times \Pi _{2}(\mathbf{M}^{2},\mathbf{s}_{0},q^{2}),  \label{eq:SRCoup}
\end{eqnarray}%
where $\Pi _{2}(\mathbf{M}^{2},\mathbf{s}_{0},q^{2})$ is the amplitude $\Pi
_{2}^{\mathrm{OPE}}(p^{2},p^{\prime 2},q^{2})$ after relevant
transformations and subtractions.

Operations to extract $g_{2}(q^{2})$ have been detailed above and are
standard recipes of the SR method. In numerical analysis, we use $m_{\eta
_{c}}=(2984.1\pm 0.4)~\mathrm{MeV}$, $m_{\eta _{b}}=(9398.7~\pm 2.0)\
\mathrm{MeV}$ as the masses of the mesons $\eta _{c}$ and $\eta _{b}$ \cite%
{PDG:2024}. We employ also the decay constants $f_{\eta _{c}}=(421\pm 35)~%
\mathrm{MeV}$, and $f_{\eta _{b}}=724~\mathrm{MeV}$, where the former is the
sum rule prediction \cite{Veliev:2010vd}. For the parameters $M_{2}^{2}$,
and $s_{0}^{\prime }$ in the $\eta _{b}$ channel, we have applied the
working windows
\begin{equation}
M_{2}^{2}\in \lbrack 10,12]~\mathrm{GeV}^{2},\ s_{0}^{\prime }\in \lbrack
95,99]~\mathrm{GeV}^{2}.
\end{equation}

The extrapolating function $\mathcal{G}_{2}(Q^{2})$ with parameters $%
\mathcal{G}_{2}^{0}=16.53~\mathrm{GeV}^{-1}$, $c_{2}^{1}=8.45$, and $%
c_{2}^{2}=-14.10$ nicely agrees with QCD predictions. The strong coupling $%
g_{2}$ amounts to
\begin{equation}
g_{2}\equiv \mathcal{G}_{2}(-m_{\eta _{c}}^{2})=(10.07\pm 1.91)\ \mathrm{GeV}%
^{-1}.
\end{equation}%
The partial width of this process is equal to
\begin{equation}
\Gamma \left[ \mathcal{M}_{\mathrm{T}}\rightarrow \eta _{b}\eta _{c}\right]
=g_{2}^{2}\frac{\lambda _{2}}{40\pi m^{2}}|M_{2}|^{2},
\end{equation}%
where%
\begin{eqnarray}
&&|M_{2}|^{2}=\frac{\left[ m^{4}+(m_{\eta _{b}}^{2}-m_{\eta
_{c}}^{2})^{2}-2m^{2}(m_{\eta _{b}}^{2}+m_{\eta _{c}}^{2})\right] ^{2}}{%
24m^{4}},  \notag \\
&&  \label{eq:M2}
\end{eqnarray}%
and $\lambda _{2}=\lambda (m,m_{\eta _{b}},m_{\eta _{c}})$.

The decay $\mathcal{M}_{\mathrm{T}}\rightarrow \eta _{b}\eta _{c}$ has the
width
\begin{equation}
\Gamma \left[ \mathcal{M}_{\mathrm{T}}\rightarrow \eta _{b}\eta _{c}\right]
=(18.3\pm 8.6)~\mathrm{MeV}.  \label{eq:DW20}
\end{equation}

%%%%%%%%%%%%%%%%%%%%%%%%%%%%%%%%%%%%%%%%%%%%%%%%%%%%%%%%%%%%%%%%%%5

\section{Decays $\mathcal{M}_{\mathrm{T}}\rightarrow B_{c}^{\ast
+}B_{c}^{\ast -}$ and $B_{c}^{+}B_{c}^{-}$}

\label{sec:Widths2}

%%%%%%%%%%%%%%%%%%%%%%%%%%%%%%%%%%%%%%%%%%%%%%%%%%%%%%%%%%%%%%%%%%%%%%%

In this section, we explore decays of the tensor molecule $\mathcal{M}_{%
\mathrm{T}}$ to $B_{c}^{\ast +}B_{c}^{\ast -}$ and $B_{c}^{+}B_{c}^{-}$
mesons. It is easy to see that the decays $\mathcal{M}_{\mathrm{T}%
}\rightarrow B_{c}^{\ast +}B_{c}^{\ast -}$ and $B_{c}^{+}B_{c}^{-}$ with
thresholds $12.68~\mathrm{GeV}$ and $12.55~\mathrm{GeV}$ are\ kinematically
allowed modes of the tensor molecule $\mathcal{M}_{\mathrm{T}}$.

For the masses of these mesons, we are going to employ%
\begin{eqnarray}
m_{B_{c}} &=&(6274.47\pm 0.27\pm 0.17)~\mathrm{MeV},  \notag \\
m_{B_{c}^{\ast }} &=&6338~\mathrm{MeV},
\end{eqnarray}%
where first of them is the experimental datum \cite{PDG:2024}, whereas the
second one is the theoretical prediction from Ref.\ \cite{Godfrey:2004ya}.
The decay constant of these mesons are
\begin{equation}
f_{B_{c}}=(371\pm 37)~\mathrm{MeV},\ f_{B_{c}^{\ast }}=471~\mathrm{MeV},
\end{equation}%
which are borrowed from Refs.\ \cite{Wang:2024fwc,Eichten:2019gig},
respectively.

%%%%%%%%%%%%%%%%%%%%%%%%%%%%%%%%%%%%%%%%%%%%%%%%%%%%%%%%%%%%%%%%%%%%%%%%

\subsection{$\mathcal{M}_{\mathrm{T}}\rightarrow B_{c}^{\ast +}B_{c}^{\ast
-} $}

%%%%%%%%%%%%%%%%%%%%%%%%%%%%%%%%%%%%%%%%%%%%%%%%%%%%%%%%%%%%%%%%%%%%%%%%%%%

Analysis of this decay is done in accordance with the scheme presented
above. Therefore, we provide only decisive formulas and predictions.

Our aim is to extract SR for the form factor $g_{3}(q^{2})$ which describe
the interaction of particles at the vertex $\mathcal{M}_{\mathrm{T}%
}B_{c}^{\ast +}B_{c}^{\ast -}$. The relevant correlator is%
\begin{eqnarray}
\widetilde{\Pi }_{\mu \nu \alpha \beta }(p,p^{\prime }) &=&i^{2}\int
d^{4}xd^{4}ye^{ip^{\prime }y}e^{-ipx}\langle 0|\mathcal{T}\{J_{\mu
}^{B_{c}^{\ast +}}(y)  \notag \\
&&\times J_{\nu }^{B_{c}^{\ast -}}(0)J_{\alpha \beta }^{\dagger
}(x)\}|0\rangle .
\end{eqnarray}%
Here, $J_{\mu }^{B_{c}^{\ast +}}$ and $J_{\nu }^{B_{c}^{\ast -}}$ are the
currents that correspond to $B_{c}^{\ast +}$ and $B_{c}^{\ast -}$ mesons
\begin{equation}
J_{\mu }^{B_{c}^{\ast +}}(x)=\overline{b}_{i}(x)\gamma _{\mu }c_{i}(x),\
J_{\nu }^{B_{c}^{\ast -}}(x)=\overline{c}_{j}(x)\gamma _{\nu }b_{j}(x).
\end{equation}

After some operations the correlation function $\widetilde{\Pi }_{\mu \nu
\alpha \beta }^{\mathrm{Phys}}(p,p^{\prime })$ acquires the form
\begin{eqnarray}
&&\widetilde{\Pi }_{\mu \nu \alpha \beta }^{\mathrm{Phys}}(p,p^{\prime })=%
\frac{\langle 0|J_{\mu }^{B_{c}^{\ast +}}|B_{c}^{\ast +}(p^{\prime
},\varepsilon _{1})\rangle }{p^{\prime 2}-m_{B_{c}^{\ast }}^{2}}\frac{%
\langle 0|J_{\nu }^{B_{c}^{\ast -}}|B_{c}^{\ast -}(q,\varepsilon
_{2})\rangle }{q^{2}-m_{B_{c}^{\ast }}^{2}}  \notag \\
&&\times \langle B_{c}^{\ast +}(p^{\prime },\varepsilon _{1})B_{c}^{\ast
-}(q,\varepsilon _{2})|\mathcal{M}_{\mathrm{T}}(p,\epsilon )\rangle \frac{%
\langle \mathcal{M}_{\mathrm{T}}(p,\epsilon )|J_{\alpha \beta }^{\dagger
}|0\rangle }{p^{2}-m^{2}}  \notag \\
&&+\cdots .
\end{eqnarray}%
Subsequent calculations are performed using the matrix elements
\begin{eqnarray}
&&\langle 0|J_{\mu }^{B_{c}^{\ast +}}|B_{c}^{\ast +}(p^{\prime },\varepsilon
_{1})\rangle =f_{B_{c}^{\ast }}m_{B_{c}^{\ast }}\varepsilon _{1\mu
}(p^{\prime }),  \notag \\
&&\langle 0|J_{\nu }^{B_{c}^{\ast -}}|B_{c}^{\ast -}(q,\varepsilon
_{2})\rangle =f_{B_{c}^{\ast }}m_{B_{c}^{\ast }}\varepsilon _{2\nu }(q),
\end{eqnarray}%
where $\varepsilon _{1\mu }(p^{\prime })$ and $\varepsilon _{2\nu }(q)$ are
the polarization vectors of $B_{c}^{\ast +}$ and $B_{c}^{\ast -}$,
respectively. The vertex $\mathcal{M}_{\mathrm{T}}B_{c}^{\ast +}B_{c}^{\ast
-}$ is treated as in Eq.\ (\ref{eq:TVV}) with substitution $%
g_{1}(q^{2})\rightarrow g_{3}(q^{2})$.

Then, the correlation function $\widetilde{\Pi }_{\mu \nu \alpha \beta }^{%
\mathrm{Phys}}(p,p^{\prime })$ in terms of the parameters of the molecule $%
\mathcal{M}_{\mathrm{T}}$ and $B_{c}^{\ast \pm }$ mesons is
\begin{eqnarray}
&&\widetilde{\Pi }_{\mu \nu \alpha \beta }^{\mathrm{Phys}}(p,p^{\prime })=%
\frac{g_{3}(q^{2})\Lambda f_{B_{c}^{\ast }}^{2}m_{B_{c}^{\ast }}^{2}}{\left(
p^{2}-m^{2}\right) \left( p^{\prime 2}-m_{B_{c}^{\ast }}^{2}\right) }  \notag
\\
&&\times \frac{1}{(q^{2}-m_{B_{c}^{\ast }}^{2})}\left[ \frac{1}{2}p_{\mu
}p_{\alpha }^{\prime }g_{\nu \beta }+\frac{m^{2}+m_{B_{c}^{\ast }}^{2}-q^{2}%
}{4m^{2}}p_{\mu }^{\prime }p_{\alpha }g_{\beta \nu }\right.  \notag \\
&&\left. +p_{\beta }^{\prime }p_{\alpha }^{\prime }g_{\mu \nu }+\text{other
structures}\right] +\cdots .
\end{eqnarray}%
The function $\widetilde{\Pi }_{\mu \nu \alpha \beta }(p,p^{\prime })$
computed using quark propagators amounts to
\begin{eqnarray}
&&\widetilde{\Pi }_{\mu \nu \alpha \beta }^{\mathrm{OPE}}(p,p^{\prime
})=i\int d^{4}xd^{4}ye^{ip^{\prime }y}e^{-ipx}\mathrm{Tr}\left[ \gamma _{\mu
}S_{c}^{ia}(y-x)\right.  \notag \\
&&\left. \times \gamma _{\beta }S_{b}^{ai}(x-y)\right] \mathrm{Tr}\left[
\gamma _{\nu }S_{b}^{jb}(-x)\gamma _{\alpha }S_{c}^{bj}(x)\right] .  \notag
\\
&&
\end{eqnarray}

The sum rule for $g_{3}(q^{2})$
\begin{eqnarray}
&&g_{3}(q^{2})=\frac{2(q^{2}-m_{B_{c}^{\ast }}^{2})}{\Lambda f_{B_{c}^{\ast
}}^{2}m_{B_{c}^{\ast }}^{2}}e^{m^{2}/M_{1}^{2}}e^{m_{B_{c}^{\ast
}}^{2}/M_{2}^{2}}  \notag \\
&&\times \Pi _{3}(\mathbf{M}^{2},\mathbf{s}_{0},q^{2})
\end{eqnarray}%
is extracted by employing the amplitudes $\Pi _{3}^{\mathrm{Phys}%
}(p^{2},p^{\prime 2},q^{2})$ and $\Pi _{3}^{\mathrm{OPE}}(p^{2},p^{\prime
2},q^{2})$ which correspond to terms $\sim p_{\mu }p_{\alpha }^{\prime
}g_{\nu \beta }$ in the correlators $\widetilde{\Pi }_{\mu \nu \alpha \beta
}^{\mathrm{Phys}}(p,p^{\prime })$ and $\widetilde{\Pi }_{\mu \nu \alpha
\beta }^{\mathrm{OPE}}(p,p^{\prime })$, respectively. Above $\Pi _{3}(%
\mathbf{M}^{2},\mathbf{s}_{0},q^{2})$ is the amplitude $\Pi _{3}^{\mathrm{OPE%
}}(p^{2},p^{\prime 2},q^{2})$ obtained after relevant transformations.

Numerical calculations have been performed by employing, in the $B_{c}^{\ast
+}$ meson's channel, the parameters $M_{2}^{2}$ and$\ s_{0}^{\prime }$ as
\begin{equation}
M_{2}^{2}\in \lbrack 6.5,7.5]~\mathrm{GeV}^{2},\ s_{0}^{\prime }\in \lbrack
49,51]~\mathrm{GeV}^{2}.
\end{equation}%
The constants of the function $\mathcal{G}_{3}(Q^{2})$ are equal to $%
\mathcal{G}_{3}^{0}=0.59~\mathrm{GeV}^{-1}$, $c_{3}^{1}=3.53$, and $%
c_{3}^{2}=-2.84$. For the coupling $g_{3}$, we obtain
\begin{equation}
g_{3}\equiv \mathcal{G}_{3}(-m_{B_{c}^{\ast }}^{2})=(2.1\pm 0.4)\times
10^{-1}\ \mathrm{GeV}^{-1}.
\end{equation}%
The partial width of the decay $\mathcal{M}_{\mathrm{T}}\rightarrow
B_{c}^{\ast +}B_{c}^{\ast -}$ is equal to
\begin{eqnarray}
&&\Gamma \left[ \mathcal{M}_{\mathrm{T}}\rightarrow B_{c}^{\ast
+}B_{c}^{\ast -}\right] =\frac{g_{3}^{2}\lambda _{3}}{80\pi m^{2}}\left(
m^{4}-3m^{2}m_{B_{c}^{\ast }}^{2}\right.  \notag \\
&&\left. +6m_{B_{c}^{\ast }}^{4}\right) ,
\end{eqnarray}%
where $\lambda _{3}=\lambda (m,m_{B_{c}^{\ast }},m_{B_{c}^{\ast }})$.

Numerical calculations yield
\begin{equation}
\Gamma \left[ \mathcal{M}_{\mathrm{T}}\rightarrow B_{c}^{\ast +}B_{c}^{\ast
-}\right] =(20.7\pm 6.9)~\mathrm{MeV}.
\end{equation}

%%%%%%%%%%%%%%%%%%%%%%%%%%%%%%%%%%%%%%%%%%%%%%%%%%%%%%%%%%%%%%%%%%%%%%%%

\subsection{$\mathcal{M}_{\mathrm{T}}\rightarrow B_{c}^{+}B_{c}^{-}$}

%%%%%%%%%%%%%%%%%%%%%%%%%%%%%%%%%%%%%%%%%%%%%%%%%%%%%%%%%%%%%%%%%%%%%%%%%%%

The decay $\mathcal{M}_{\mathrm{T}}\rightarrow B_{c}^{+}B_{c}^{-}$ is
studied in the same way. We consider the correlation function
\begin{eqnarray}
\widetilde{\Pi }_{\mu \nu }(p,p^{\prime }) &=&i^{2}\int
d^{4}xd^{4}ye^{ip^{\prime }y}e^{-ipx}\langle 0|\mathcal{T}\{\
J^{B_{c}^{+}}(y)  \notag \\
&&\times J^{B_{c}^{-}}(0)J_{\mu \nu }^{\dagger }(x)\}|0\rangle ,
\end{eqnarray}%
with $\ J^{B_{c}^{+}}(x)$ and $J^{B_{c}^{-}}(x)$ being the interpolating
currents of the $B_{c}^{+}$ and $B_{c}^{-}$ mesons
\begin{equation}
\ J^{B_{c}^{+}}(x)=\overline{b}_{i}(x)i\gamma _{5}c_{i}(x),\
J^{B_{c}^{-}}(x)=\overline{c}_{j}(x)i\gamma _{5}b_{j}(x).
\end{equation}

The matrix elements of the pseudoscalar $B_{c}^{\pm }$ mesons are
\begin{equation}
\langle 0|J^{B_{c}^{\pm }}|B_{c}^{\pm }\rangle =\frac{f_{B_{c}}m_{B_{c}}^{2}%
}{m_{b}+m_{c}}.
\end{equation}%
The vertex $\langle B_{c}^{+}(p^{\prime })B_{c}^{-}(q)|T(p,\epsilon )\rangle
$ has the form
\begin{equation}
\langle B_{c}^{+}(p^{\prime })B_{c}^{-}(q)|\mathcal{M}_{\mathrm{T}%
}(p,\epsilon )\rangle =g_{4}(q^{2})\epsilon _{\alpha \beta }^{(\lambda
)}(p)p^{\prime \alpha }p^{\prime \beta }.
\end{equation}%
The $\widetilde{\Pi }_{\mu \nu }^{\mathrm{Phys}}(p,p^{\prime })$ is given
after some substitutions by Eq.\ (\ref{eq:CF4}), while Eq.\ (\ref{eq:CF5})
determines the QCD side of SR. The form factor $g_{4}(q^{2})$ is
\begin{eqnarray}
&&g_{4}(q^{2})=\frac{(m_{b}+m_{c})^{2}(q^{2}-m_{B_{c}}^{2})}{\Lambda
f_{B_{c}}^{2}m_{B_{c}}^{4}}e^{m^{2}/M_{1}^{2}}e^{m_{B_{c}}^{2}/M_{2}^{2}}
\notag \\
&&\times \Pi _{4}(\mathbf{M}^{2},\mathbf{s}_{0},q^{2}).
\end{eqnarray}

In calculations, we have utilized the parameters
\begin{equation}
M_{2}^{2}\in \lbrack 6.5,7.5]~\mathrm{GeV}^{2},\ s_{0}^{\prime }\in \lbrack
45,47]~\mathrm{GeV}^{2}.
\end{equation}%
Computations of the form factor $g_{4}(q^{2})$ and coupling $g_{4}$ lead to
the result
\begin{equation}
g_{4}\equiv \mathcal{G}_{4}(-m_{B_{c}}^{2})=(10.67\pm 1.92)\ \mathrm{GeV}%
^{-1},
\end{equation}%
where $\mathcal{G}_{4}(Q^{2})$ is the fitting function with parameters $%
\mathcal{G}_{4}^{0}=14.62~\mathrm{GeV}^{-1}$, $c_{4}^{1}=3.09$, and $%
c_{4}^{2}=7.42$.

The width of the decay $\mathcal{M}_{\mathrm{T}}\rightarrow
B_{c}^{+}B_{c}^{-}$ is calculated by utilizing the formula
\begin{eqnarray}
\Gamma \left[ \mathcal{M}_{\mathrm{T}}\rightarrow B_{c}^{+}B_{c}^{-}\right]
&=&g_{4}^{2}\frac{\lambda _{4}}{960\pi m^{2}}\left(
m^{2}-4m_{B_{c}}^{2}\right) ^{2},  \notag \\
&&  \label{eq:PDw3}
\end{eqnarray}%
where $\lambda _{4}=\lambda (m,m_{B_{c}},m_{B_{c}})$. This formula can be
obtained from Eq.\ (\ref{eq:M2}) in the limit $m_{\eta _{b}}$ and $m_{\eta
_{c}}\rightarrow m_{B_{c}}$ . Our computations yield
\begin{equation}
\Gamma \left[ \mathcal{M}_{\mathrm{T}}\rightarrow B_{c}^{+}B_{c}^{-}\right]
=(21.2\pm 12.3)~\mathrm{MeV}.
\end{equation}

%%%%%%%%%%%%%%%%%%%%%%%%%%%%%%%%%%%%%%%%%%%%%%%%%%%%%%%%%%%%%%%%%

\section{Decays generated by $b\overline{b}$ annihilations}

\label{sec:Widths3}

%%%%%%%%%%%%%%%%%%%%%%%%%%%%%%%%%%%%%%%%%%%%%%%%%%%%%%%%%%%

The hadronic molecule $\mathcal{M}_{\mathrm{T}}$ can be converted to
ordinary mesons also after annihilation of $b\overline{b}$ quarks to light
quark-antiquark pairs \cite{Becchi:2020mjz,Becchi:2020uvq,Agaev:2023ara} and
generation of $DD$ mesons of appropriate parameters. We study here the
decays of the molecule $\mathcal{M}_{\mathrm{T}}$ to $D^{+}D^{-}$, $D^{0}%
\overline{D}^{0}$, $D^{\ast +}D^{\ast -}$, $D^{\ast 0}\overline{D}^{\ast 0}$%
, $D_{s}^{+}D_{s}^{-}$ and $D_{s}^{\ast +}D_{s}^{\ast -}$ mesons.

These modes are kinematically possible channels of the molecule $\mathcal{M}%
_{\mathrm{T}}$ to conventional particles. We investigate these decays also
in the three-point SR method. Here, to calculate the correlation functions
which contain$\ $vacuum matrix element $\langle \overline{b}b\rangle $ of $b$
quark, we replace it with the known gluon condensate $\langle \alpha
_{s}G^{2}/\pi \rangle $ \cite{Agaev:2023ara}. But implementation of
annihilation processes into the three-point SR framework does not imply
usage of additional free parameters.

%%%%%%%%%%%%%%%%%%%%%%%%%%%%%%%%%%%%%%%%%%%%%%%%%%%%%%%%%%%%%%%%%%%%%%%%%%%%

\subsection{Decays $\mathcal{M}_{\mathrm{T}}\rightarrow D^{\ast 0}\overline{D%
}^{\ast 0}$ and $D^{\ast +}D^{\ast -}$}

%%%%%%%%%%%%%%%%%%%%%%%%%%%%%%%%%%%%%%%%%%%%%%%%%%%%%%%%%%%%%%%%%%%%%%%%%%%%

First, we analyze the process $\mathcal{M}_{\mathrm{T}}\rightarrow D^{\ast 0}%
\overline{D}^{\ast 0}$. To find the coupling $G_{1}$ of particles at the
vertex $\mathcal{M}_{\mathrm{T}}D^{\ast 0}\overline{D}^{\ast 0}$, we
consider the correlator%
\begin{eqnarray}
\widehat{\Pi }_{\mu \nu \alpha \beta }(p,p^{\prime }) &=&i^{2}\int
d^{4}xd^{4}ye^{ip^{\prime }y}e^{-ipx}\langle 0|\mathcal{T}\{J_{\mu }^{%
\overline{D}^{\ast 0}}(y)  \notag \\
&&\times J_{\nu }^{D^{\ast 0}}(0)J_{\alpha \beta }^{\dagger }(x)\}|0\rangle ,
\label{eq:CF1A}
\end{eqnarray}%
with $J_{\mu }^{\overline{D}^{\ast 0}}(x)$ and $J_{\nu }^{D^{\ast 0}}(x)$
being the interpolating currents for the mesons $\overline{D}^{\ast 0}$ and $%
D^{\ast 0}$
\begin{equation}
J_{\mu }^{\overline{D}^{\ast 0}}(x)=\overline{c}_{i}(x)\gamma _{\mu
}u_{i}(x),\text{ }J_{\nu }^{D^{\ast 0}}(x)=\overline{u}_{j}(x)\gamma _{\nu
}c_{j}(x).  \label{eq:CRB}
\end{equation}

\begin{figure}[h]
\includegraphics[width=8.5cm]{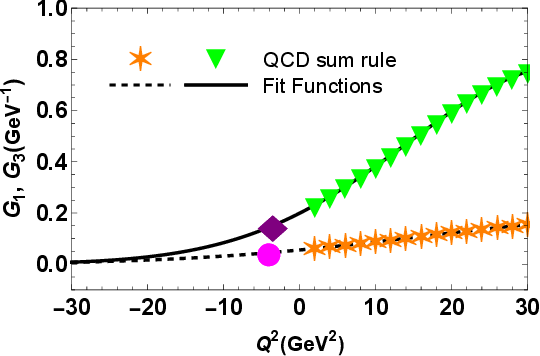}
\caption{QCD data and extrapolating functions for the form factors $%
G_{1}(Q^{2})$ (dashed curve) and $G_{3}(Q^{2})$ (solid curve). The circle
and diamond fix the points $Q^{2}=-m_{D^{\ast 0}}^{2}$ and $%
Q^{2}=-m_{D^{0}}^{2}$, respectively. }
\label{fig:Fit1}
\end{figure}

The correlation function $\widehat{\Pi }_{\mu \nu \alpha \beta }(p,p^{\prime
})$ in terms of the matrix elements of the particles $\mathcal{M}_{\mathrm{T}%
}$, $D^{\ast 0}$, and $\overline{D}^{\ast 0}$ is
\begin{eqnarray}
&&\widehat{\Pi }_{\mu \nu \alpha \beta }^{\mathrm{Phys}}(p,p^{\prime })=%
\frac{\langle 0|J_{\mu }^{\overline{D}^{\ast 0}}|\overline{D}^{\ast
0}(p^{\prime },\varepsilon _{1})\rangle }{p^{\prime 2}-m_{D^{\ast 0}}^{2}}%
\frac{\langle 0|J_{\nu }^{D^{\ast 0}}|D^{\ast 0}(q,\varepsilon _{2})\rangle
}{q^{2}-m_{D^{\ast 0}}^{2}}  \notag  \label{eq:CF2} \\
&&\times \langle \overline{D}^{\ast 0}(p^{\prime },\varepsilon _{1})D^{\ast
0}(q,\varepsilon _{2})|\mathcal{M}_{\mathrm{T}}(p,\epsilon )\rangle \frac{%
\langle \mathcal{M}_{\mathrm{T}}(p,\epsilon )|J_{\mu }^{\dagger }|0\rangle }{%
p^{2}-m^{2}}  \notag \\
&&+\cdots ,
\end{eqnarray}%
where $m_{D^{\ast 0}}=(2006.85\pm 0.05)~\mathrm{MeV}$ is the mass of the
mesons $\overline{D}^{\ast 0}$ and $D^{\ast 0}$, whereas $\varepsilon _{1\mu
}$ and $\varepsilon _{2\nu }$ are their polarization vectors.

The function $\widehat{\Pi }_{\mu \nu \alpha \beta }^{\mathrm{Phys}%
}(p,p^{\prime })$ is found by means of the matrix elements
\begin{eqnarray}
\langle 0|J_{\mu }^{\overline{D}^{\ast 0}}|\overline{D}^{\ast 0}(p^{\prime
},\varepsilon _{1})\rangle &=&f_{D^{\ast }}m_{D^{\ast 0}}\varepsilon _{1\mu
}(p^{\prime }),  \notag \\
\langle 0|J_{\nu }^{D^{\ast 0}}|D^{\ast 0}(q,\varepsilon _{2})\rangle
&=&f_{D^{\ast }}m_{D^{\ast 0}}\varepsilon _{2\nu }(q),  \label{eq:ME2B}
\end{eqnarray}%
with $f_{D^{\ast }}=(252.2\pm 22.66)~\mathrm{MeV}$ being the decay constant
of the mesons $D^{\ast 0}$ and $\overline{D}_{D}^{\ast 0}$. The vertex $%
\langle \overline{D}^{\ast 0}(p^{\prime },\varepsilon _{1})D^{\ast
0}(q,\varepsilon _{2})|\mathcal{M}_{\mathrm{T}}(p,\epsilon )\rangle $ is
written down in the form of Eq.\ (\ref{eq:TVV}).

The correlation function $\widehat{\Pi }_{\mu \nu \alpha \beta }^{\mathrm{%
Phys}}(p,p^{\prime })$ contains different components. The sum rule for the
form factor $G_{1}(q^{2})$ is obtained by employing the amplitude $\widehat{%
\Pi }_{1}^{\mathrm{Phys}}(p^{2},p^{\prime 2},q^{2})$ corresponding to the
term $\sim p_{\mu }p_{\alpha }^{\prime }g_{\beta \nu }$. The correlator $%
\widehat{\Pi }_{\mu \nu \alpha \beta }(p,p^{\prime })$ calculated by
employing the quark propagators amounts to
\begin{eqnarray}
&&\widehat{\Pi }_{\mu \nu \alpha \beta }^{\mathrm{OPE}}(p,p^{\prime })=\frac{%
1}{3}\langle \overline{b}b\rangle \int d^{4}xd^{4}ye^{ip^{\prime }y}e^{-ipx}
\notag \\
&&\times \mathrm{Tr}\left[ \gamma _{\mu }S_{c}^{ja}(y-x)\gamma _{\beta
}\gamma _{\alpha }S_{c}^{ai}(x)\gamma _{\nu }{}S_{u}^{ij}(-y)\right] ,
\label{eq:QCDsideA}
\end{eqnarray}%
where $S_{u}(x)$ is the $u$ quark's propagator \cite{Agaev:2020zad}. We
denote by $\widehat{\Pi }_{1}^{\mathrm{OPE}}(p^{2},p^{\prime 2},q^{2})$ the
amplitude that in $\widehat{\Pi }_{\mu \nu \alpha \beta }^{\mathrm{OPE}%
}(p,p^{\prime })$ corresponds to the same term $\sim p_{\mu }p_{\alpha
}^{\prime }g_{\beta \nu }$.

To continue computations, we utilize the relation between the condensates
\begin{equation}
\langle \overline{b}b\rangle =-\frac{1}{12m_{b}}\langle \frac{\alpha
_{s}G^{2}}{\pi }\rangle  \label{eq:Conden}
\end{equation}%
obtained in Ref.\ \cite{Shifman:1978bx} by employing the sum rules.

The sum rule for the form factor $G_{1}(q^{2})$ is%
\begin{eqnarray}
&&G_{1}(q^{2})=\frac{2(q^{2}-m_{D^{\ast 0}}^{2})}{\Lambda f_{D^{\ast
}}^{2}m_{D^{\ast 0}}^{2}}e^{m^{2}/M_{1}^{2}}e^{m_{D^{\ast 0}}^{2}/M_{2}^{2}}
\notag \\
&&\times \widehat{\Pi }_{1}(\mathbf{M}^{2},\mathbf{s}_{0},q^{2}),
\end{eqnarray}%
where $\widehat{\Pi }_{1}(\mathbf{M}^{2},\mathbf{s}_{0},q^{2})$ is the
amplitude $\widehat{\Pi }_{1}^{\mathrm{OPE}}(p^{2},p^{\prime 2},q^{2})$
after the Borel transformations and continuum subtractions.

In numerical computations for the $\overline{D}^{\ast 0}$ meson channel, we
utilize the parameters%
\begin{equation}
M_{2}^{2}\in \lbrack 2,3]~\mathrm{GeV}^{2},\ s_{0}^{\prime }\in \lbrack
5.7,5.8]~\mathrm{GeV}^{2}.  \label{eq:Wind2}
\end{equation}%
To evaluate the coupling $G_{1}$ we use the SR data for $Q^{2}=2-30\ \mathrm{%
GeV}^{2}$ and the extrapolating function with parameters $\widehat{\mathcal{G%
}}_{1}^{0}=0.056~\mathrm{GeV}^{-1}$, $\widehat{c}_{1}^{1}=8.32$, and $%
\widehat{c}_{1}^{2}=-14.87$. The sum rule data and function $\widehat{%
\mathcal{G}}_{1}(Q^{2})$ are plotted in Fig.\ \ref{fig:Fit1}. The coupling $%
G_{1}$ is calculated at $q^{2}=m_{D^{\ast 0}}^{2}$
\begin{equation}
G_{1}\equiv \widehat{\mathcal{G}}_{1}(-m_{D^{\ast 0}}^{2})=(4.5\pm
0.9)\times 10^{-2}\ \mathrm{GeV}^{-1}.  \label{eq:G1}
\end{equation}%
The width of the decay $\mathcal{M}_{\mathrm{T}}\rightarrow D^{\ast 0}%
\overline{D}^{\ast 0}$ is
\begin{equation}
\Gamma \left[ \mathcal{M}_{\mathrm{T}}\rightarrow D^{\ast 0}\overline{D}%
^{\ast 0}\right] =(7.63\pm 2.17)~\mathrm{MeV}.
\end{equation}

The difference between the second $\mathcal{M}_{\mathrm{T}}\rightarrow
D^{\ast +}D^{\ast -}$ and first decays is encoded in the masses of the
final-state mesons. The QCD side of the channel $\mathcal{M}_{\mathrm{T}%
}\rightarrow D^{\ast +}D^{\ast -}$ can be obtained from Eq.\ (\ref%
{eq:QCDsideA}) by the substitution $S_{u}^{ij}(-y)\rightarrow S_{d}^{ij}(-y)$%
. Because in this paper we adopt the equality $m_{u}=m_{d}=0$ this means
that QCD sides of these two decays are identical to each other. Therefore,
with nice accuracy $\Gamma \left[ \mathcal{M}_{\mathrm{T}}\rightarrow
D^{\ast +}D^{\ast -}\right] \approx \Gamma \left[ \mathcal{M}_{\mathrm{T}%
}\rightarrow D^{\ast 0}\overline{D}^{\ast 0}\right] $. Nevertheless, we have
carried out corresponding computations and found
\begin{equation}
\Gamma \left[ \mathcal{M}_{\mathrm{T}}\rightarrow D^{\ast +}D^{\ast -}\right]
=(7.67\pm 2.17)~\mathrm{MeV}.
\end{equation}%
As it has been expected, the difference between partial widths of these
channels is very small.

%%%%%%%%%%%%%%%%%%%%%%%%%%%%%%%%%%%%%%%%%%%%%%%%%%%%%%%%%%%%%%%%%%%%%%%%%%%%

\subsection{Processes $\mathcal{M}_{\mathrm{T}}\rightarrow D^{0}\overline{D}%
^{0}$ and $D^{+}D^{-}$}

%%%%%%%%%%%%%%%%%%%%%%%%%%%%%%%%%%%%%%%%%%%%%%%%%%%%%%%%%%%%%%%%%%%%%%%%%%%%

The channel $\mathcal{M}_{\mathrm{T}}\rightarrow D^{0}\overline{D}^{0}$ is
considered using the correlation function
\begin{eqnarray}
\widehat{\Pi }_{\mu \nu }(p,p^{\prime }) &=&i^{2}\int
d^{4}xd^{4}ye^{ip^{\prime }y}e^{-ipx}\langle 0|\mathcal{T}\{J^{\overline{D}%
^{0}}(y)  \notag \\
&&\times J^{D^{0}}(0)J_{\mu \nu }^{\dagger }(x)\}|0\rangle ,
\end{eqnarray}%
where the currents $J^{\overline{D}^{0}}(x)$ and $J^{D^{0}}(x)$ are
introduced by formulas%
\begin{equation}
J^{\overline{D}^{0}}(x)=\overline{c}_{i}(x)i\gamma _{5}u_{i}(x),\text{ }%
J^{D^{0}}(x)=\overline{u}_{j}(x)i\gamma _{5}c_{j}(x).
\end{equation}%
To derive the sum rule for the form factor $G_{3}(q^{2})$ that describes the
strong interaction of particles at the vertex $\mathcal{M}_{\mathrm{T}}D^{0}%
\overline{D}^{0}$, we have to find the correlators $\widehat{\Pi }_{\mu \nu
}^{\mathrm{Phys}}(p,p^{\prime })$ and $\widehat{\Pi }_{\mu \nu }^{\mathrm{OPE%
}}(p,p^{\prime })$.

We determine $\widehat{\Pi }_{\mu \nu }^{\mathrm{Phys}}(p,p^{\prime })$ by
means of the matrix elements
\begin{equation}
\langle 0|J^{\overline{D}^{0}}|\overline{D}^{0}\rangle =\langle
0|J^{D^{0}}|D^{0}\rangle =\frac{f_{D}m_{D^{0}}^{2}}{m_{c}},
\end{equation}%
and
\begin{equation}
\langle \overline{D}^{0}(p^{\prime })D^{0}(q)|T(p,\epsilon )\rangle
=G_{3}(q^{2})\epsilon _{\alpha \beta }^{(\lambda )}(p)p^{\prime \alpha
}p^{\prime \beta },
\end{equation}%
where $m_{D^{0}}=(1864.84\pm 0.05)~\mathrm{MeV}$ and $f_{D}=(211.9\pm 1.1)~%
\mathrm{MeV}$ are the mass and decay constant of mesons $D^{0}$ and $%
\overline{D}^{0}$ \cite{PDG:2024,Rosner:2015wva}. As a result, we find
\begin{eqnarray}
&&\widehat{\Pi }_{\mu \nu }^{\mathrm{Phys}}(p,p^{\prime })=\frac{%
G_{3}(q^{2})\Lambda f_{D}^{2}m_{D^{0}}^{4}}{m_{c}^{2}\left(
p^{2}-m^{2}\right) \left( p^{\prime 2}-m_{D^{0}}^{2}\right) \left(
q^{2}-m_{D^{0}}^{2}\right) }  \notag \\
&&\times \left[ \frac{%
m^{4}-2m^{2}(m_{D^{0}}^{2}+q^{2})+(m_{D^{0}}^{2}-q^{2})^{2}}{12m^{2}}g_{\mu
\nu }\right.  \notag \\
&&\left. +p_{\mu }^{\prime }p_{\nu }^{\prime }-\frac{%
m^{2}+m_{D^{0}}^{2}-q^{2}}{2m^{2}}p_{\mu }p_{\nu }^{\prime }+\text{other
terms}\right] .
\end{eqnarray}%
For $\widehat{\Pi }_{\mu \nu }^{\mathrm{OPE}}(p,p^{\prime })$, we get%
\begin{eqnarray}
&&\widehat{\Pi }_{\mu \nu }^{\mathrm{OPE}}(p,p^{\prime })=\frac{1}{3}\langle
\overline{b}b\rangle \int d^{4}xd^{4}ye^{ip^{\prime }y}e^{-ipx}  \notag \\
&&\times \mathrm{Tr}\left[ \gamma _{5}{}S_{c}^{ja}(y-x)\gamma _{\mu }\gamma
_{\nu }S_{c}^{ai}(x)\gamma _{5}S_{u}^{ij}(-y)\right] .  \label{eq:QCDsideB}
\end{eqnarray}%
To obtain the sum rule for $G_{3}(q^{2})$, we employ the amplitudes $%
\widehat{\Pi }_{3}^{\mathrm{Phys}}(p^{2},p^{\prime 2},q^{2})$ and $\widehat{%
\Pi }_{3}^{\mathrm{OPE}}(p^{2},p^{\prime 2},q^{2})$ corresponding to the
structure $p_{\mu }^{\prime }p_{\nu }^{\prime }$ and find
\begin{eqnarray}
&&G_{3}(q^{2})=\frac{m_{c}^{2}(q^{2}-m_{D^{0}}^{2})}{\Lambda
f_{D}^{2}m_{D^{0}}^{4}}e^{m^{2}/M_{1}^{2}}e^{m_{D^{0}}^{2}/M_{2}^{2}}  \notag
\\
&&\times \widehat{\Pi }_{3}(\mathbf{M}^{2},\mathbf{s}_{0},q^{2}),
\end{eqnarray}%
with $\widehat{\Pi }_{3}(\mathbf{M}^{2},\mathbf{s}_{0},q^{2})$ being the
transformed function $\widehat{\Pi }_{3}^{\mathrm{OPE}}(p^{2},p^{\prime
2},q^{2})$.

In numerical calculations we have used the parameters
\begin{equation}
M_{2}^{2}\in \lbrack 1.5,3]~\mathrm{GeV}^{2},\ s_{0}^{\prime }\in \lbrack
5,5.2]~\mathrm{GeV}^{2}.
\end{equation}%
The coupling $G_{3}$ is estimated by means of the function $\widehat{%
\mathcal{G}}_{3}(Q^{2})$ with $\widehat{\mathcal{G}}_{3}^{0}=0.20~\mathrm{GeV%
}^{-1}$, $\widehat{c}_{3}^{1}=11.02$, and $\widehat{c}_{3}^{2}=-28.26$
\begin{equation}
G_{3}\equiv \widehat{\mathcal{G}}_{3}(-m_{D^{0}}^{2})=(1.53\pm 0.28)\times
10^{-1}\ \mathrm{GeV}^{-1}.
\end{equation}%
The partial width of the decay $\mathcal{M}_{\mathrm{T}}\rightarrow D^{0}%
\overline{D}^{0}$ is equal to
\begin{equation}
\Gamma \left[ \mathcal{M}_{\mathrm{T}}\rightarrow D^{0}\overline{D}^{0}%
\right] =(6.64\pm 1.73)~\mathrm{MeV}.
\end{equation}%
The width of the process $\mathcal{M}_{\mathrm{T}}\rightarrow D^{+}D^{-}$,
as it has been explained above, is approximately equal to $\Gamma \left[
\mathcal{M}_{\mathrm{T}}\rightarrow D^{0}\overline{D}^{0}\right] $. Indeed,
our computations yield
\begin{equation}
\Gamma \left[ \mathcal{M}_{\mathrm{T}}\rightarrow D^{+}D^{-}\right]
=(6.45\pm 1.64)~\mathrm{MeV}.
\end{equation}

%%%%%%%%%%%%%%%%%%%%%%%%%%%%%%%%%%%%%%%%%%%%%%%%%%%%%%%%%%%%%%%%%%%%%%%%%%%%

\subsection{Channels $\mathcal{M}_{\mathrm{T}}\rightarrow D_{s}^{+}D_{s}^{-}$
and $D_{s}^{\ast +}D_{s}^{\ast -}$}

%%%%%%%%%%%%%%%%%%%%%%%%%%%%%%%%%%%%%%%%%%%%%%%%%%%%%%%%%%%%%%%%%%%%%%%%%%%%

Investigation of these channels does not differ considerably from analysis
of the modes considered in the previous subsections. There are only a few
replacements in the correlation functions and parameters of the mesons which
should be taken into account.

Thus, the correlators of the decays $\mathcal{M}_{\mathrm{T}}\rightarrow
D_{s}^{\ast +}D_{s}^{\ast -}$ and $\mathcal{M}_{\mathrm{T}}\rightarrow
D_{s}^{+}D_{s}^{-}$ are obtained from Eqs.\ (\ref{eq:QCDsideA}) and (\ref%
{eq:QCDsideB}) after replacing the propagators $S_{u}^{ij}(-y)$ by $%
S_{s}^{ij}(-y)$. In numerical computations we include in analysis terms $%
\sim m_{s}=(93.5\pm 0.8)~\mathrm{MeV}$, but ignore ones proportional to $%
m_{s}^{2}$. It is clear that the masses and decay constants of the $D_{s}$
and $D_{s}^{\ast }$ mesons
\begin{eqnarray}
m_{D_{s}} &=&(1969.0\pm 1.4)~\mathrm{MeV},\ f_{D_{s}}=(249.9\pm 0.5)~\mathrm{%
MeV},  \notag \\
&&
\end{eqnarray}%
and
\begin{eqnarray}
m_{D_{s}^{\ast }} &=&(2112.2\pm 0.4)~\mathrm{MeV},\ f_{D_{s}^{\ast
}}=(268.8\pm 6.5)~\mathrm{MeV},  \notag \\
&&
\end{eqnarray}%
appear in calculations as input parameters.

The widths of these decays are
\begin{equation}
\Gamma \left[ \mathcal{M}_{\mathrm{T}}\rightarrow D_{s}^{+}D_{s}^{-}\right]
=(6.34\pm 1.72)~\mathrm{MeV},
\end{equation}%
and
\begin{equation}
\Gamma \left[ \mathcal{M}_{\mathrm{T}}\rightarrow D_{s}^{\ast +}D_{s}^{\ast
-}\right] =(7.38\pm 2.10)~\mathrm{MeV},
\end{equation}%
respectively.

%%%%%%%%%%%%%%%%%%%%%%%%%%%%%%%%%%%%%%%%%%%%%%%%%%%%%%%%%%%%%%%%%

\section{Decays into $B$ meson pairs}

\label{sec:Widths4}

%%%%%%%%%%%%%%%%%%%%%%%%%%%%%%%%%%%%%%%%%%%%%%%%%%%%%%%%%%

Decays of the hadronic molecule $\mathcal{M}_{\mathrm{T}}$ into pairs of the
$B$ and $B_{s}$ mesons are investigated by the same manner. These decays are
generated by $c\overline{c}$ annihilation which
lead to creations of $BB$ and $B_{s}B_{s}$ mesons with correct charges and
quantum numbers. It is easy to see that $\mathcal{M}_{\mathrm{T}}\rightarrow
B^{(\ast )+}B^{(\ast )-}$, $B^{(\ast )0}\overline{B}^{(\ast )0}$, and $%
B_{s}^{0}\overline{B}_{s}^{0}$ are allowed decay channels of the hadronic
molecule $\mathcal{M}_{\mathrm{T}}$.

Below, as a sample, we consider the decay $\mathcal{M}_{\mathrm{T}%
}\rightarrow B^{\ast +}B^{\ast -}$. The correlation function necessary to
evaluate the coupling $\overline{G}_{1}$ at the vertex $\mathcal{M}_{\mathrm{%
T}}B^{\ast +}B^{\ast -}$ is
\begin{eqnarray}
\overline{\Pi }_{\mu \nu \alpha \beta }(p,p^{\prime }) &=&i^{2}\int
d^{4}xd^{4}ye^{ip^{\prime }y}e^{-ipx}\langle 0|\mathcal{T}\{J_{\mu
}^{B^{\ast +}}(y)  \notag \\
&&\times J_{\nu }^{B^{\ast -}}(0)J_{\alpha \beta }^{\dagger }(x)\}|0\rangle ,
\end{eqnarray}%
where $J^{B^{\ast +}}(x)$ and $J^{B^{\ast -}}(x)$ are the interpolating
currents of the mesons $B^{\ast +}$ and $B^{\ast -}$
\begin{equation}
J_{\mu }^{B^{\ast +}}(x)=\overline{b}_{i}(x)\gamma _{\mu }u_{i}(x),\ J_{\nu
}^{B^{\ast -}}(x)=\overline{u}_{j}(x)\gamma _{\nu }b_{j}(x).
\end{equation}%
In calculation of the correlator $\overline{\Pi }_{\mu \nu \alpha \beta }^{%
\mathrm{Phys}}(p,p^{\prime })$, we have used the matrix elements
\begin{eqnarray}
&&\langle 0|J_{\mu }^{B^{\ast +}}|B^{\ast +}(p^{\prime },\varepsilon
_{1})\rangle =f_{B^{\ast }}m_{B^{\ast }}\varepsilon _{1\mu }(p^{\prime }),
\notag \\
&&\langle 0|J_{\nu }^{B^{\ast -}}|B^{\ast -}(q,\varepsilon _{2})\rangle
=f_{B^{\ast }}m_{B^{\ast }}\varepsilon _{2\nu }(q).
\end{eqnarray}%
In these expressions $m_{B^{\ast }}=(5324.75\pm 0.20)~\mathrm{MeV}$ and $%
f_{B^{\ast }}=(210\pm 6)~\mathrm{MeV}$ are the mass and decay constant of
the $B^{\ast \pm }$ mesons. The matrix element of the vertex $\mathcal{M}_{%
\mathrm{T}}B^{\ast +}B^{\ast -}$ is given by Eq.\ (\ref{eq:TVV}) after
evident replacements.

The correlator $\overline{\Pi }_{\mu \nu \alpha \beta }(p,p^{\prime })$
through quark propagators can be expressed by the following formula
\begin{eqnarray}
&&\overline{\Pi }_{\mu \nu \alpha \beta }^{\mathrm{OPE}}(p,p^{\prime })=%
\frac{1}{3}\langle \overline{c}c\rangle \int d^{4}xd^{4}ye^{ip^{\prime
}y}e^{-ipx}  \notag \\
&&\times \mathrm{Tr}\left[ \gamma _{\mu }S_{u}^{ij}(y)\gamma _{\nu
}S_{b}^{ja}(-x)\gamma _{\alpha }\gamma _{\beta }S_{b}^{ai}(x-y){}\right] .
\end{eqnarray}%
The condensate $\langle \overline{c}c\rangle $ with some accuracy can be
replaced by \cite{Shifman:1978bx}
\begin{equation}
\langle \overline{c}c\rangle =-\frac{1}{12m_{c}}\langle \frac{\alpha
_{s}G^{2}}{\pi }\rangle .
\end{equation}

Numerical calculations of the form factor $\overline{G}_{1}(q^{2})$ are
performed for $Q^{2}=2-30~\mathrm{GeV}^{2}$. The Borel and continuum
subtraction parameters in the $B^{\ast +}$ meson channel are varied within
limits%
\begin{equation}
M_{2}^{2}\in \lbrack 5.5,6.5]~\mathrm{GeV}^{2},\ s_{0}^{\prime }\in \lbrack
34,35]~\mathrm{GeV}^{2}.
\end{equation}%
The coupling $\overline{G}_{1}$ is extracted at $Q^{2}=-m_{B^{\ast }}^{2}$
by utilizing the fit function $\overline{\mathcal{G}}_{1}(Q^{2})$ with
parameters $\overline{\mathcal{G}}_{1}^{0}=0.104~\mathrm{GeV}^{-1}$, $%
\overline{c}_{1}^{1}=2.71$, and $\overline{c}_{1}^{2}=-3.63$
\begin{equation}
\overline{G}_{1}\equiv \overline{\mathcal{G}}_{1}(-m_{B^{\ast
}}^{2})=(5.89\pm 1.12)\times 10^{-2}\ \mathrm{GeV}^{-1}.
\end{equation}%
Then the width of the process $\mathcal{M}_{\mathrm{T}}\rightarrow B^{\ast
+}B^{\ast -}$ is equal to
\begin{equation}
\Gamma \left[ \mathcal{M}_{\mathrm{T}}\rightarrow B^{\ast +}B^{\ast -}\right]
=(5.46\pm 1.48)~\mathrm{MeV}.
\end{equation}

Another decays of $\mathcal{M}_{\mathrm{T}}$ to $B$ and $B_{s}$ mesons are
studied in this manner. Results of relevant analyses are collected in Table %
\ref{tab:Channels}, in which we give information on strong couplings and
partial widths of these processes. Note that parameters of the decays $%
\mathcal{M}_{\mathrm{T}}\rightarrow B^{(\ast )0}\overline{B}^{(\ast )0}$ are
chosen equal to those of $\mathcal{M}_{\mathrm{T}}\rightarrow B^{(\ast
)+}B^{(\ast )-}$ and omitted in the Table.

Information on the partial widths of the channels considered in last three
sections permits us to estimate the full width of the hadronic molecule
\begin{equation}
\Gamma \left[ \mathcal{M}_{\mathrm{T}}\right] =(154\pm 19)~\mathrm{MeV}.
\end{equation}

\begin{table}[tbp]
\begin{tabular}{|c|c|c|c|}
\hline\hline
i & Channels & $\overline{G}_{i}~(\mathrm{GeV}^{-1})$ & $\Gamma_{i}~(\mathrm{%
MeV})$ \\ \hline
$1$ & $B^{\ast +}B^{\ast -}$ & $(5.89 \pm 1.12)\times 10^{-2}$ & $5.46 \pm
1.48$ \\
$2$ & $B^{+}B^{-}$ & $0.48 \pm 0.09$ & $4.98 \pm 1.38$ \\
$3$ & $B_{s}^{\ast 0}\overline{B}_{s}^{\ast 0}$ & $(5.07 \pm 1.01)\times
10^{-2}$ & $3.86 \pm 1.10 $ \\
$4$ & $B_{s}^{0}\overline{B}_{s}^{0}$ & $0.44 \pm 0.08$ & $3.47 \pm 0.94 $
\\ \hline\hline
\end{tabular}%
\caption{Decay channels of the hadronic molecule $\mathcal{M}_{\mathrm{T}}$
to $B$ and $B_s$ mesons, related couplings $\overline{G}_{i}$ and widths $%
\Gamma_{i}$.}
\label{tab:Channels}
\end{table}

\section{Discussion and conclusions}

\label{sec:Conc}

In this article, we have computed important parameters of the hadronic
tensor molecule $\mathcal{M}_{\mathrm{T}}=B_{c}^{\ast +}B_{c}^{\ast -}$ in
the context of QCD sum rule method: We have evaluated its mass and full
decay width. The mass of $\mathcal{M}_{\mathrm{T}}$ has been calculated by
employing the two-point sum rule approach, whereas decays of these molecule
have been investigated using technical tools of the three-point SR method.

In our previous publications \cite{Agaev:2025wdj,Agaev:2025fwm}, we explored
parameters of the scalar and axial-vector molecules built of $B_{c}$ mesons.
Comparing prediction for the masses of these states with $m=(12.87\pm 0.08)~%
\mathrm{GeV}$, we see that this particle is heavier than its counterparts.
But this difference is relatively small and amounts to $\sim 100~\mathrm{MeV}
$. It is interesting also to confront the tensor molecule $\mathcal{M}_{%
\mathrm{T}}$ with the tensor diquark-antidiquark state $T$ \ with the same
quark content \cite{Agaev:2024qbh}. The mass of the tetraquark $T$ was found
there to be equal to $(12.70\pm 0.09)~\mathrm{GeV}$. As is seen, the gap
between masses of these two structures is small and does not exceed $170~%
\mathrm{MeV}$, and even has a common point at $12.79~\mathrm{GeV}$ provided
one takes into account uncertainties of computations.

The molecules $B_{c}^{(\ast )+}B_{c}^{(\ast )-}$ were also considered in the
framework of the coupled-channel unitary approach \cite{Liu:2023gla}. In
this article some integrals were regularized by including the parameter $%
\Lambda $. As a result, predictions for the masses of the molecules depend
on a choice of $\Lambda $. For the mass of the tensor molecule $B_{c}^{\ast
+}B_{c}^{\ast -}$ the maximal value $12660.7~\mathrm{MeV}$ was obtained at $%
\Lambda =400~\mathrm{MeV}$, whereas the minimal value $12287.7~\mathrm{MeV}$
was found at $\Lambda =1400~\mathrm{MeV}$. Theoretical ambiguities of these
results are approximately $\sim 400~\mathrm{MeV}$. Even maximal value of the
mass is smaller than our result. Note that the similar picture was observed
while considering the scalar and axial-vector molecules as well.

Our prediction $m=(12.87\pm 0.08)~\mathrm{GeV}$ proves that it is unstable
against the strong decays to different pairs of conventional mesons. The
processes $\mathcal{M}_{\mathrm{T}}\rightarrow J/\psi \Upsilon $, $\eta
_{b}\eta _{c}$ and $B_{c}^{(\ast )+}B_{c}^{(\ast )-}$ are main fall-apart
channels of the molecule $\mathcal{M}_{\mathrm{T}}$: In these decays all
heavy quarks from $\mathcal{M}_{\mathrm{T}}$ appear in the final-state
mesons. Alternative modes of $\mathcal{M}_{\mathrm{T}}$ are due to
annihilations of $b\overline{b}$ and $c\overline{c}$ quarks inside of the
molecule and generation of various processes. Note that contribution of
these decays to the full width of the molecule $\mathcal{M}_{\mathrm{T}}$
constitutes approximately a half of this parameter. Our result for $\Gamma %
\left[ \mathcal{M}_{\mathrm{T}}\right] $ demonstrate that $\mathcal{M}_{%
\mathrm{T}}$ is a wide molecular structure, which can decay through two
different mechanisms.

Four-quark mesons composed of heavy quarks are an import component of the
exotic hadron spectroscopy. Hadronic molecules $B_{c}^{(\ast )+}B_{c}^{(\ast
)-}$ or diquark-antidiquark states with the same content $bc\overline{b}%
\overline{c}$ were not discovered, but they may be observed in the future
LHC and FCC experiments \cite{Carvalho:2015nqf,Abreu:2023wwg}. Therefore,
detailed theoretical studies of such systems by means of different
approaches and comparison of the obtained predictions are necessary to
prepare useful ground for these experiments.

\end{document}